\documentclass[a4paper,11pt]{article}
\pdfoutput=1
\usepackage{jheppub}
\usepackage[utf8]{inputenc}
\usepackage{mathtools}

\usepackage{tikz}
\usetikzlibrary{positioning}

\renewcommand{\d}{\mathrm{d}}

\tikzset{
smallnode/.style={circle, draw, very thick, minimum size=2mm},
smallcircle/.style={circle, fill, scale=0.5},
roundnode/.style={circle, draw, very thick, minimum size=10mm},
squarednode/.style={rectangle, draw, very thick, minimum size=10mm}}
\tikzset{every loop/.style={}}

\preprint{DESY 18-046}

\title{\boldmath Gauge theories from principally extended \\ disconnected gauge groups}

\author{Antoine Bourget$^a$, Alessandro Pini$^b$ and Diego Rodríguez-Gómez$^a$}

\affiliation{${}^a$ Department of Physics, Universidad de Oviedo,\\ C/Federico García Lorca 18, 33007 Oviedo, Spain\\
${}^b$ DESY Theory Group Notkestraße 85, 22607 Hamburg, Germany
}
\emailAdd{bourgetantoine@uniovi.es, alessandro.pini@desy.de, \\ d.rodriguez.gomez@uniovi.es}

\abstract{We introduce gauge theories based on a class of disconnected gauge groups, called principal extensions. Although in this work we focus on 4d theories with $\mathcal{N}=2$ SUSY, such construction is independent of spacetime dimensions and supersymmetry. These groups implement in a consistent way the discrete gauging of charge conjugation, for arbitrary rank. Focusing on the principal extension of $\mathrm{SU}(N)$, we explain how many of the exact methods for theories with 8 supercharges can be put into practice in that context. We then explore the physical consequences of having a disconnected gauge group: we find that the Coulomb branch is generically non-freely generated, and the global symmetry of the Higgs branch is modified in a non-trivial way.}

\begin{document} 
\maketitle
\flushbottom

\section{Introduction and Conclusions}

Gauge theories lie at the core of modern Physics. As such, a huge body of work has been devoted to their study. Nevertheless, it is fair to say that most of it starts by assuming a connected gauge group. In turn, at least comparatively, very little attention has been paid to theories based on disconnected gauge groups, although some early works exist \cite{krauss1989discrete,banks1989effective} (see also \cite{Banks:2010zn,Gaiotto:2014kfa} for a more recent revival). Very recently, there has been, however, a spark of activity along these lines \cite{Argyres:2016yzz,Argyres:2017tmj} (see also \cite{Caorsi:2018ahl,Xie:2017obm,Agarwal:2017roi,Bourget:2017tmt,Argyres:2015ffa,Argyres:2015gha,Argyres:2016xmc,McOrist:2013bga}), motivated by the quest to find more generic gauge theories in 4d with $\mathcal{N}=2$ supersymmetry. Also, disconnected gauge groups, obtained by gauging discrete symmetries, play also a relevant role in constructing the recently discovered strictly $\mathcal{N}=3$ SUSY \cite{Garcia-Etxebarria:2015wns} -- see also \cite{Garcia-Etxebarria:2017ffg,Aharony:2016kai,Lemos:2016xke,Argyres:2018wxu,Tom}. In this paper we will study a particular class of disconnected groups well-known in the mathematical literature but somewhat exotic in the physics context\footnote{To our knowledge, principal extensions have made a brief appearance in the physical literature --e.g. \cite{Bachas:2000ik,Maldacena:2001xj,Stanciu:2001vw}--, albeit in the different context of branes on group manifolds.}, namely \textit{principal extensions} of a classical group counterpart. To set the notations, in the following we will denote by $\widetilde{G}$ the principal extension of a classical ``parent" group $G$. The principal extension group $\widetilde{G}$ is obtained by taking a semidirect group of $G$ with the automorphism group $\Gamma$ of its Dynkin diagram. 

Even though this construction may sound contrived, secretly, we are already familiar with it, since the good old $\mathrm{O}(2N)$ groups can be regarded as the principal extension of $\mathrm{SO}(2N)$ (see  \cite{Bourget:2017tmt} for a recent discussion in the Physics context). In that case, the Dynkin diagram is symmetric under the exchange of the two nodes in the tail, and the non-trivial element of $\Gamma$ is this exchange. In the case of $\mathrm{SU}(N)$ the Dynkin diagram is symmetric under flipping, which is in this case the non-trivial element of $\Gamma$. Note that the operation of flipping the Dynkin diagram is equivalent to exchanging the fundamental and the antifundamental of $G$, which clearly reminds of charge conjugation. Actually, a similar interpretation holds for the $\mathrm{O}(2N)$ case. Hence, the group $\widetilde{G}$ may be interpreted as a version of $G$ where charge conjugation has been gauged. Therefore principal extensions implement \textit{ab initio} the gauging of charge conjugation (note that the semidirect product structure is important to this matter) in an automatically consistent way and for any rank. 

It is important to stress that these principal extensions are, at the end of the day, just Lie groups. Thus, from this perspective, we can come back to Physics and construct gauge theories based on them. Note that this is very generic, independent of spacetime dimension and, of course of supersymmetry. Nevertheless, and for the sake of definiteness, in the following, we will concentrate on the case of 4d gauge theories with $\mathcal{N}=2$ supersymmetry, for which very powerful exact techniques are available, allowing us to study the theories in detail. It is worth to stress, in this context, that our approach, based on an intrinsically consistent mathematical structure, allows to construct an integration formula with which to easily compute exact partition functions \cite{wendt2001weyl}. 

Since ultimately the principal extensions are nothing but Lie groups, the construction of gauge theories based on them simply follows from the standard rules (as in, for instance, \cite{Tachikawa:2013kta}). In our case, since we will study SCQD-like theories, the basic building blocks are $\mathcal{N}=2$ vector multiplets in the adjoint of $\widetilde{G}$ and a number $N_f$ of hypermultiplets in the fundamental of $\widetilde{G}$. The difference with the standard theory based on the parent $G$ group is that in the matter and interactions in principal extension theory come in representations of $\widetilde{G}$. Perhaps this is best illustrated looking to the matter sector. While in the theory based on $G$ we have a hypermultiplet in the fundamental representation $\mathbf{N}$ of $G$, in the theory based on $\widetilde{G}$ we will have matter in a real representation constructed as $\mathbf{N}+\bar{\mathbf{N}}$. Note that while this is not an irreducible representation of $G$, it is the fundamental irreducible of $\widetilde{G}$, and it implements in a direct way the invariance under the exchange of $\mathbf{N}$ and $\bar{\mathbf{N}}$.

Another consequence of the fact that $\widetilde{G}$ is just a Lie group whose connected part is $G$ is that, when quantizing the theory based on $\widetilde{G}$, we will obtain the same Feynman rules as in the theory based on $G$. In the end the reason is that, perturbatively, fields are close to the identity, and thus only sensitive to the connected part of $\widetilde{G}$ which is identical to $G$.\footnote{That the local dynamics is unchanged after gauging discrete symmetries has been already emphasized in \cite{Aharony:2016kai,Argyres:2016yzz}.} Hence, in particular, the conditions for conformal invariance are identical to those in the theory based on $G$. That means, for the particular case which will mostly concern us in this paper of $\widetilde{\mathrm{SU}}(N)$, that we should supply $2N$ flavors. 

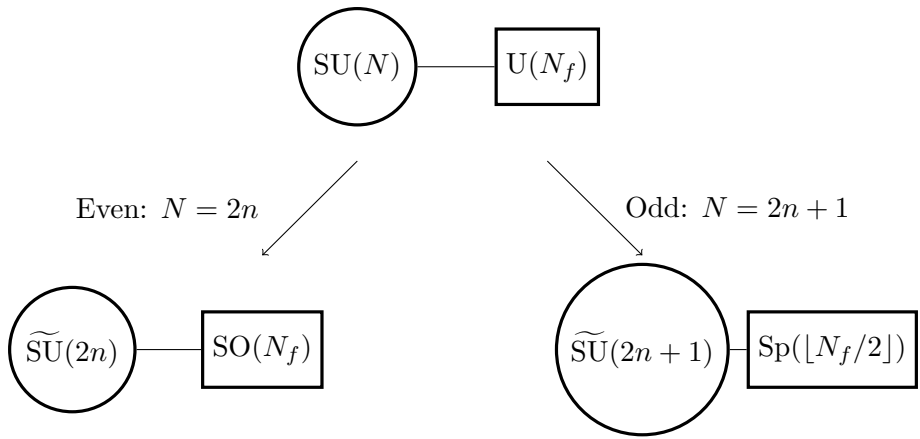
\begin{figure}[t]
    \centering
\def\x{2.5}
\begin{tikzpicture}
\node[roundnode] (0) at (0*\x,0*\x) {$\mathrm{SU}(N)$};
\node[squarednode] (1) at (1*\x,0*\x) {$\mathrm{U}(N_f)$};
\draw[-] (0) -- (1);
\node[roundnode] (10) at (-1.5*\x,-1.5*\x) {$\widetilde{\mathrm{SU}}(2n)$};
\node[squarednode] (11) at (-0.5*\x,-1.5*\x) {$\mathrm{SO}(N_f)$};
\draw[-] (10) -- (11);
\node[roundnode] (20) at (1.5*\x,-1.5*\x) {$\widetilde{\mathrm{SU}}(2n+1)$};
\node[squarednode] (21) at (2.5*\x,-1.5*\x) {$\mathrm{Sp}(\lfloor N_f/2\rfloor)$};
\draw[-] (20) -- (21);
\draw [->] (0*\x,-0.5*\x) -- (-0.5*\x,-1*\x);
\draw [->] (1*\x,-0.5*\x) -- (1.5*\x,-1*\x);
\node (3) at (-1*\x,-0.75*\x) {Even: $N=2n$};
\node (4) at (2*\x,-0.75*\x) {Odd: $N=2n+1$};
\end{tikzpicture}
    \caption{Summary of the global symmetries of QCD with gauge group $\mathrm{SU}(N)$ and $\widetilde{\mathrm{SU}}(N)$ and $N_f$ flavors. }
    \label{figSummaryGlobalSymQCD}
\end{figure}

Having said this, one may then ask in what respect the theory based on $\widetilde{G}$ is different to the theory based on $G$. As we will see below, this difference strikingly manifests itself in the spectrum of operators of the theory. Since the gauge group has been enlarged, the set of gauge-invariants should be expected to be smaller. As we will see, this has spectacular consequences, such as, in particular, a typically non-freely generated Coulomb branch (see \cite{Argyres:2017tmj} for a general discussion). Other striking property of the theories is that, mirroring the fact that the fundamental of $\widetilde{\mathrm{SU}}(N)$ is real, the global symmetry of QCD must have a real or pseudo-real fundamental representation. This rules out the traditional $\mathrm{U}(N_f)$, and it turns out that the global symmetry for $\widetilde{\mathrm{SU}}(2N)$ is $\mathrm{SO}(N_f)$, while the global symmetry for $\widetilde{\mathrm{SU}}(2N+1)$ is $\mathrm{Sp}(\lfloor\frac{N_f}{2}\rfloor)$, as represented schematically in Figure \ref{figSummaryGlobalSymQCD}. 

This work, in a sense a proof of concept, only touches upon the tip of an iceberg. Gauge theories based on principal extensions can be defined in arbitrary dimensions and with no mention to supersymmetry. Thus, it would be very interesting to survey the landscape of these theories in general. Restricting to the particularly controlled 4d $\mathcal{N}=2$ set-up described in this paper, we have only studied the most direct aspects of $\widetilde{\mathrm{SU}}(N)$ theories, leaving many things open for further studies. To begin with, it is compelling to study the global aspects associated to $\widetilde{\mathrm{SU}}(N)$. Having disconnected components, one may imagine that the $\widetilde{\mathrm{SU}}(N)$ theory depends on discrete parameters controlling the relative weight of the different sectors (this discussion may heavily depend on the spacetime dimension). A related issue, very sensitive to the global aspects of the gauge group (e.g. \cite{Aharony:2013hda}) is that of extended operators -- such as line and surface defects --, whose classification would be extremely interesting to study (in particular in view of \cite{Gaiotto:2014kfa}). One striking feature of the $\widetilde{\mathrm{SU}}(N)$ theories is the fact that they have a non-freely generated Coulomb branch. Hence, it would be captivating to construct the Seiberg-Witten curve and study its properties, extending the results in \cite{Argyres:2017tmj}. It would also be very interesting to study these theories with ``exotic" Higgs branches and non-freely generated Coulomb branches using chiral algebras and bootstrap as in \cite{Beem:2014rza,Lemos:2015awa}. Once we have these new class of theories based on $\widetilde{\mathrm{SU}}(N)$, a natural further step is to combine them into quiver theories -- our analysis of global symmetries suggest in particular the existence of linear quivers with alternating $\widetilde{\mathrm{SU}}(N)$ groups and orthogonal/symplectic groups, depending on the parity of $N$. We postpone the study of these quivers --and in general, of the landscape of these new theories along the lines of \cite{Bhardwaj:2013qia}-- to future work, with special attention to the conditions under which the quiver remains conformal. It would also be interesting to explore their large $N$ limit, paying special attention to their putative gravity dual.

A natural further question is the stringy embedding of these theories, and in particular whether they fit into the class $\mathcal{S}$ scheme. In this respect, it would be very interesting to clarify the relation to the construction of twisted gauge theories on $S^1\times \mathcal{M}_3$ introduced in \cite{Zwiebel:2011wa} and extended in \cite{Mekareeya:2012tn} to the general case. Those papers consider a twisted version of the original $SU(N)$ gauge theory by the action, as one goes around an $S^1$ in the background geometry, of the same outer automorphism as in our case.\footnote{Actually, to be more precise, with a version of it --charge conjugation-- identical to ours up to conjugation by a $SU(N)$ element --see eq.\eqref{equationConjugation}.} Hence their partition functions are very related to the contribution of the sector disconnected with the identity in our case. Indeed, the result for the vector multiplet is identical to ours (for instance one may compare the integrands in eqs. (3.15), (3.17) in \cite{Mekareeya:2012tn} --which extend those in \cite{Zwiebel:2011wa}-- with that in our eq. \eqref{eq:cindex}). Yet, the contribution of hypermultiplets (eq. (4.42) in \cite{Mekareeya:2012tn}) does not coincide with ours in the denominator of eq.\eqref{HiggsBranchFormula}, the latter being actually identical to eq. (4.42) in \cite{Mekareeya:2012tn} upon setting $u=1$. We stress however that while refs. \cite{Zwiebel:2011wa,Mekareeya:2012tn} twist the original theory by an action of the automorphism group, we are constructing a theory based on a new --disconnected-- group (and therefore must include both the connected and disconnected pieces).

The rest of this paper is organized as follows. In section \ref{sectionGtilde} we introduce, mainly from a mathematical point of view, the principal extension groups. As anticipated, we will mostly concentrate on the case of $\mathrm{SU}(N)$ and its principal extension $\widetilde{\mathrm{SU}}(N)$. In particular we will study some aspects of its representations and invariant theory, with special focus on the adjoint and the fundamental representations. Based solely on group theory, we can already see at this level that the ring of invariants of the group is not freely generated, which translates, in Physics language, to a non-freely generated Coulomb branch. In section \ref{sectionGtilde} we will also introduce the integration formula, which will be very useful when we turn to the computation of exact partition functions for the associated gauge theories. In section \ref{SUSYtheories} we turn into Physics by sketching the construction of $\mathcal{N}=2$ SUSY theories based on $\widetilde{\mathrm{SU}}(N)$. As briefly discussed above, we focus on $\mathcal{N}=2$ 4d SUSY theories just for definiteness, but we could consider re-evaluating any gauge theory in arbitrary dimensions by basing it on $\widetilde{G}$. In section \ref{CB} we study in detail the Coulomb branch of the theory by computing the Coulomb branch limit of the $\mathcal{N}=2$ index, finding in particular that it is not freely generated. In section \ref{sectionHiggs} we turn to the Higgs branch and analyze the global symmetry by looking to the Higgs branch Hilbert series -- which we expect to coincide with the Hall-Littlewood limit of the 4d $\mathcal{N}=2$ index. We postpone in the appendices several technical details, including a discussion of the ring of invariants of the $\widetilde{\mathrm{SU}}(N)$ theory and its non-freely generated property from a purely mathematical point of view. We also compute the Coulomb branch index for a theory based on the principal extension of $\widetilde{E}_6$, also leading to a non-freely generated Coulomb branch.

\section{\texorpdfstring{The group $\widetilde{G}$}{The group G tilde}}
\label{sectionGtilde}

In this section, we describe a family of Lie groups called the \emph{principal extensions} \cite{wendt2001weyl}, and describe some of their properties that are used in the rest of the paper. In short, the principal extension of a connected and simply connected Lie group $G$ is a disconnected group $\widetilde{G}$ whose connected component is $G$ and whose group of connected components $\widetilde{G}/G$ is isomorphic to the group of automorphisms of the Dynkin diagram of $G$. 

\begin{figure}[t]
    \centering
      \begin{tikzpicture}
        \node (0) at (-2.5,0) {$A_{N-1}$};
\node[smallnode] (1) at (0,0) {};
\node[smallnode] (2) at (1,0) {};
\node (3) at (2,0) {$\cdots$};
\node[smallnode] (4) at (3,0) {};
\node[smallnode] (5) at (4,0) {};
\draw[-] (1) -- (2);
\draw[-] (2) -- (3);
\draw[-] (3) -- (4);
\draw[-] (4) -- (5);
\draw[<->] (0,-.3) to [bend right] node[below] {$\mathcal{P}$} (4,-.3);
\draw[<->] (1,-.3) to [bend right] node[below] {} (3,-.3);
\end{tikzpicture} 

  \begin{tikzpicture}
  \node (0) at (-2,0) {$D_N$};
\node[smallnode] (1) at (0,0) {};
\node[smallnode] (2) at (1,0) {};
\node (3) at (2,0) {$\cdots$};
\node[smallnode] (4) at (3,0) {};
\node[smallnode] (5) at (4,1) {};
\node[smallnode] (6) at (4,-1) {};
\draw[-] (1) -- (2);
\draw[-] (2) -- (3);
\draw[-] (3) -- (4);
\draw[-] (4) -- (5);
\draw[-] (4) -- (6);
\draw[<->] (4.4,.8) to [bend left] node[right] {$\mathcal{P}$} (4.4,-.8);
\end{tikzpicture} 

      \begin{tikzpicture}
            \node (0) at (-2.5,0) {$E_6$};
\node[smallnode] (1) at (0,0) {};
\node[smallnode] (2) at (1,0) {};
\node[smallnode] (3) at (2,0) {};
\node[smallnode] (13) at (2,1) {};
\node[smallnode] (4) at (3,0) {};
\node[smallnode] (5) at (4,0) {};
\draw[-] (1) -- (2);
\draw[-] (2) -- (3);
\draw[-] (3) -- (4);
\draw[-] (4) -- (5);
\draw[-] (3) -- (13);
\draw[<->] (0,-.3) to [bend right] node[below] {$\mathcal{P}$} (4,-.3);
\draw[<->] (1,-.3) to [bend right] node[below] {} (3,-.3);
\end{tikzpicture} 
    \caption{Dynkin diagram automorphisms.}
    \label{FigDynkinDiags}
\end{figure}
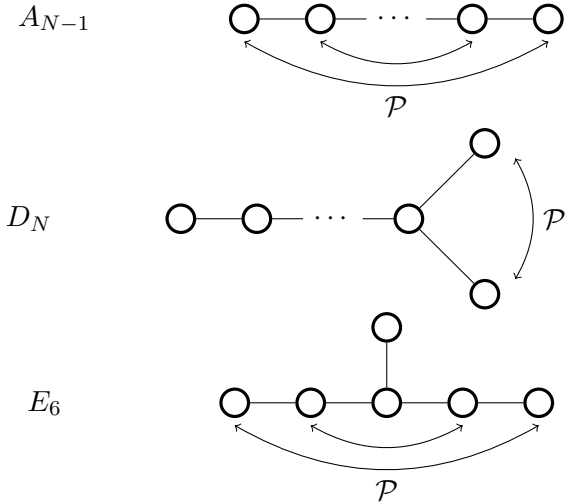

Because of this definition the cases of interest, where $\widetilde{G} \neq G$, correspond to Dynkin diagrams with non-trivial automorphisms. Three types of diagrams fall in this category (see Figure \ref{FigDynkinDiags}): 
\begin{itemize}
    \item $A_{N-1}$. The principal extension of $\mathrm{SU}(N)$ will be denoted $\widetilde{\mathrm{SU}}(N)$. This is the main focus of this article. 
    \item $D_{N}$.\footnote{Except for $N=4$, where the automorphism group permutes the three external nodes. However in this case we consider only the extension by a $\mathbb{Z}_2$ inside the group of permutations of three objects $\mathfrak{S}_3$. } In that case, the principal extension of $\mathrm{SO}(2N)$ is $\mathrm{O}(2N)$, a fact that we used in a previous paper \cite{Bourget:2017sxr}. Note that we have $\widetilde{\mathrm{SU}}(4) = \mathrm{O}(6)$. 
    \item $E_6$. In order to simplify the discussion, we will not mention this group, but most of what we say for the $A$ case has an equivalent for $E_6$. See also Appendix \ref{E6theory}. 
\end{itemize}

\subsection{\texorpdfstring{The group $\widetilde{\mathrm{SU}}(N)$}{The group \mathrm{SU}(N) tilde}}

Consider the $A_{N-1}$ Dynkin diagram, with $N \geq 3$. The associated compact, connected and simply connected Lie group is $\mathrm{SU}(N)$, and its algebra is 
\begin{equation}
\label{definitionsuN}
    \mathfrak{su}(N) = \{M \in \mathfrak{gl}(N,\mathbb{C}) | M = M^{\dagger} \textrm{ and } \mathrm{Tr} \, M = 0\} \, , 
\end{equation}
where the dagger means conjugate transpose. 

We will call $\mathcal{P}$ the non-trivial automorphism (see Figure \ref{FigDynkinDiags}, and see Appendix \ref{AppendixTechnical} for an explicit realization). The group of diagram automorphisms is then $\Gamma = \{1 , \mathcal{P}\}$. It can be shown that $\mathcal{P}$ extends uniquely to a Lie algebra automorphism of $\mathfrak{su}(N)$, and to a unique group automorphism of $\mathrm{SU}(N)$,\footnote{Here we used the fact that if $G$, $G'$ are Lie groups with corresponding Lie algebras $\mathfrak{g}$, $\mathfrak{g}'$, and with $G$ connected and simply connected, then for every Lie algebra homomorphism $\varphi : \mathfrak{g} \rightarrow \mathfrak{g}'$ there exists a unique Lie group homomorphism $\Phi : G \rightarrow G'$ such that $\varphi = \d \Phi$. See \cite{knapp2013lie}, section I.10.  } and we use the same notation for all these morphisms. 
In other words, we have a homomorphism $\varphi : \Gamma \rightarrow \mathrm{Aut}(\mathrm{SU}(N))$. Then we define the semi-direct product
\begin{equation}
\label{definitionSUtilde}
    \widetilde{\mathrm{SU}}(N) = \mathrm{SU}(N) \rtimes_{\varphi} \Gamma \, .   
\end{equation}
This means that $\widetilde{\mathrm{SU}}(N)$ is, as a set, the Cartesian product $\mathrm{SU}(N) \times \Gamma$, where the product law is given, following the standard definition of a semi-direct product, by \cite{knapp2013lie}
\begin{equation}
\label{definitionProductSUtilde}
    (X,\epsilon) \cdot (Y,\eta)  :=(X   Y^{\epsilon} ,\epsilon \eta )   \, , 
\end{equation}
where $X,Y \in \mathrm{SU}(N)$ and $\epsilon , \eta \in \Gamma$. We have used the notation 
\begin{equation}
    Y^{\epsilon} =  \varphi(\epsilon) (Y) \, . 
\end{equation}
Before going further, let us mention that the definition (\ref{definitionSUtilde}) is independent of which outer automorphism has been used in $\Gamma$ (all outer automorphisms are related by conjugation). One could have used the complex conjugation in $\mathrm{SU}(N)$ as an outer automorphism, but it will be convenient for practical computations to stick to the flipping of the Dynkin diagram $\mathcal{P}$ described here; for more about the relation between these concepts, see Appendix \ref{AppendixTechnical}. 

There are two kinds of elements, those of the form $(X,1)$ and those of the form $(X, \mathcal{P})$, belonging to the two connected components of $\widetilde{\mathrm{SU}}(N)$. Elements of the form $(X,1)$ constitute the subgroup $\mathrm{SU}(N)$ of $\widetilde{\mathrm{SU}}(N)$, while elements of the form $(X, \mathcal{P})$ make for the so-called \emph{disconnected component}. Using (\ref{definitionProductSUtilde}), one can check that the inverse of an element of $\widetilde{\mathrm{SU}}(N)$ is given by 
\begin{equation}
    (X , \epsilon)^{-1} =  ((X^{-1})^{\epsilon} , \epsilon)  \, ,  
\end{equation}
where in the right hand side we have used the fact that $\epsilon^{-1}=\epsilon$. 
Then, the group adjoint action in $\widetilde{\mathrm{SU}}(N)$ is computed as 
\begin{equation}
    (X , \epsilon) \cdot (Y , \eta) \cdot (X , \epsilon)^{-1} = (X Y^{\epsilon} (X^{-1})^{\eta} , \eta) \, . 
\end{equation}
In particular, for the particular element $(X,\epsilon) = (1, \mathcal{P}) = : \mathcal{P} $ we have 
\begin{equation}
     \mathcal{P}  \cdot  (Y , \eta)  \cdot  \mathcal{P}^{-1} = (  Y^{\mathcal{P}} , \eta ) \, . 
\end{equation}
From this, one can deduce the adjoint action of $\mathcal{P}$ on the algebra $\mathfrak{su}(N)$. As could be predicted, it is given simply by the action of\footnote{Note that by definition, $\left( e^{ix}\right)^{\mathcal{P}} = e^{i (x^{\mathcal{P}})}$ where on the left-hand side $\mathcal{P}$ acts in the group $\mathrm{SU}(N)$, and in the right-hand side in the algebra $\mathfrak{su}(N)$. } $\varphi(\mathcal{P})$,
\begin{equation}
\label{adjointActionP}
    \mathcal{P}  \cdot  x  \cdot  \mathcal{P}^{-1} = \varphi(\mathcal{P})(x) = : x^{\mathcal{P}} \, . 
\end{equation}

\subsubsection{Representations and invariants}
\label{sectionRepandInv}

In this section, we describe the representations of $\widetilde{\mathrm{SU}}(N)$ that we will use in the rest of the paper. First remark that any representation of $\widetilde{G}$ is a representation of $G$, but the converse is not true. Moreover an irreducible representation of $\widetilde{G}$ may correspond to a reducible representation of $G$. We also sketch the theory of invariants in those representations (for more details on invariant theory, see Appendix \ref{AppendixInvariant}). 

\subsubsection*{Adjoint representation}

The adjoint representation of $\widetilde{\mathrm{SU}}(N)$ is simply the representation on its Lie algebra $\mathfrak{su}(N)$, deduced from (\ref{adjointActionP}). Namely, for $\phi \in \mathfrak{su}(N)$ and $(X,\epsilon) \in \widetilde{\mathrm{SU}}(N)$, 
\begin{equation}
\label{adjointRepresentation}
    \Phi_{\mathrm{Adj}}(X,\epsilon) (\phi) = (X,\epsilon) \phi (X,\epsilon)^{-1} = X \phi^{\epsilon} X^{-1} \, . 
\end{equation}
It has dimension $N^2-1$. One can show (see equation (\ref{proofTr}) in the Appendix) that 
\begin{equation}
\label{AdjointP}
    \mathrm{Tr} ( (\phi^{\mathcal{P}}) ^k ) = (-1)^k  \mathrm{Tr} ( \phi^k ) \, . 
\end{equation}

The invariant ring of $\mathrm{SU}(N)$ in the adjoint representation is well-known, it is the freely-generated polynomial ring $\mathbb{C}[\mathrm{Tr} ( \phi^2) , \dots , \mathrm{Tr} ( \phi^N) ]$. In the case of $\widetilde{\mathrm{SU}}(N)$, the situation is more complicated, because of (\ref{AdjointP}): the $ \mathrm{Tr} ( \phi^k)$ for $k$ odd are not invariant under $\widetilde{\mathrm{SU}}(N)$. It turns out that the invariant ring is generated by the following elements: 
\begin{enumerate}
    \item[1. ] The \emph{primary} invariants $I_k$ for $2 \leq k \leq N$ defined by 
    \begin{equation}
        I_k = \begin{cases}
        \mathrm{Tr} ( \phi^k) & \textrm{ for } k \textrm{ even} \\
        \mathrm{Tr} ( \phi^k)^2 & \textrm{ for } k \textrm{ odd}  
        \end{cases} \, .
    \end{equation}
    \item[2. ] The \emph{secondary} invariants 
     \begin{equation}
        J_{k_1 , \dots , k_r} = \prod\limits_{i=1}^r \mathrm{Tr} ( \phi^{k_i}) \, ,
    \end{equation}   
    for $k_1 , \dots , k_r$ odd and $3 \leq k_1 < \dots < k_r \leq N$, with $r$ even ($r=0$ corresponds to the trivial invariant $1$).
\end{enumerate}
We refer to Appendix \ref{AppendixInvariant} for a proof of this statement. Note that as soon as non-trivial secondary invariants are present, i.e as soon as $N \geq 5$, these generators are not algebraically independent since we have, among other relations,  
\begin{equation}
\label{relationsJI}
    (J_{k,l})^2 - I_k I_l = 0 \,  , 
\end{equation}
for $k,l$ odd and $3 \leq k < l \leq N$.\footnote{For $N=5,6$ there is only one relation $(J_{3,5})^2 = I_3 I_5$. The ring is not freely generated, but it is still a complete intersection (the plethystic logarithm of the Hilbert series is a polynomial, but some coefficients are negative). For higher $N$, we don't have complete intersections anymore. For instance, for $N=7$ the relations are the three relations of the form (\ref{relationsJI}), and there are three additional relations 
\begin{equation}
    J_{3,5} J_{3,7} = I_3 J_{5,7} \, , \qquad 
       J_{3,5} J_{5,7} = I_5 J_{3,7} \, , \qquad 
          J_{3,7} J_{5,7} = I_7 J_{3,5} \, .
\end{equation}} We conclude that \emph{for $N \geq 5$, the invariant ring of $\widetilde{\mathrm{SU}}(N)$ in the adjoint representation is not freely generated.} We will rephrase this statement in more physical terms in section \ref{sec:Coulomb}.

\subsubsection*{Bifundamental representation}

The bifundamental representation has dimension $2N$. If $x,y \in \mathbb{C}^{N}$ this representation is given by\footnote{Note that, being $\widetilde{SU}(N)$ a disconnected group, in addition to the representations we construct, there may be other representations for a given highest weight.} 
\begin{equation}
\label{bifundamentalRepresentation1}
    \Phi_{\mathrm{F \bar{F}}}(X,1)  \left( \begin{array}{c} x \\ y \end{array} \right) = \left( \begin{array}{cc} X & 0 \\ 0 & \bar{X} \end{array} \right) \left( \begin{array}{c} x \\ y \end{array} \right) = \left( \begin{array}{c} X x \\ \bar{X} y \end{array} \right)  \, 
\end{equation}
and 
\begin{equation}
\label{bifundamentalRepresentation2}
    \Phi_{\mathrm{F \bar{F}}}(1,\mathcal{P})  \left( \begin{array}{c} x \\ y \end{array} \right) =  \left( \begin{array}{cc} 0 & A \\ A^{-1} & 0 \end{array} \right) \left( \begin{array}{c} x \\ y \end{array} \right) = \left( \begin{array}{c} A y \\ A^{-1} x \end{array} \right)  \, , 
\end{equation}
where $A$ is the matrix (\ref{matrixA}) such that for all $X \in \mathrm{SU}(N)$, applying the exponential map to (\ref{equationConjugation}) gives
\begin{equation}
\label{propertyA1}
    \overline{X} = A^{-1} X^{\mathcal{P}} A \, . 
\end{equation}
Note that the matrix $A$ satisfies 
\begin{equation}
\label{propertyA2}
    A^T = (-1)^{N-1} A \qquad \textrm{and} \qquad \det A = 1 \, .  
\end{equation}
This representation is irreducible. This is a crucial difference between the groups $\mathrm{SU}(N)$ and $\widetilde{\mathrm{SU}}(N)$, since in the former one would only have \eqref{bifundamentalRepresentation1}, and thus just the direct product of $\mathbf{N}$ and $\bar{\mathbf{N}}$. In Physics terms this would be the representation of a hypermultiplet in $\mathcal{N}=2$ SQCD, which, in the case of $SU(N)$, in $\mathcal{N}=1$ language, contains a chiral multiplet $Q$ in the fundamental and another chiral multiplet $\tilde{Q}$ in the antifundamental. Instead, in $\widetilde{SU}(N)$, the hypermultiplet is in a real representation $\mathbf{N}+\bar{\mathbf{N}}$, which, as discussed, is now irreducible.

Let us move on to the invariants one can build from one bifundamental. For $\mathrm{SU}(N)$, it is well-known that the invariant ring is $\mathbb{C}[y^T x]$.\footnote{Of course, here $N_f=1$ in the language of the next paragraph, and we have $y^T x = x^T y = \frac{1}{2}(x^T y   + y^T x)$. } When going to $\widetilde{\mathrm{SU}}(N)$, one has to compute the action of $\mathcal{P}$ on $y^T x$: 
\begin{equation}
\label{mesonTransfoP}
    y^T x \rightarrow  x^T  (A^{-1})^T A y = (-1)^{N-1} y^T x \, . 
\end{equation}
This is invariant only if $N$ is odd. When $N$ is even, the square of this quantity is invariant, and one can show that the invariant ring is exactly $\mathbb{C}[y^T x]$ in the first case, and $\mathbb{C}[(y^T x)^2]$ in the second case. See Appendix \ref{sectionInvBifund} for the details. 

\subsubsection*{$N_f$ Times the Bifundamental representation}

Finally, we consider the direct sum of $N_f$ times the bifundamental representation, say $x_1 , \dots , x_{N_f}$ and $y_1 , \dots , y_{N_f}$, using the notations of the previous paragraph. The $\mathrm{SU}(N)$ invariants one can construct are well known, and fall in two categories. Borrowing the terminology from Physics: 
\begin{itemize}
    \item The ``mesons" $y_i^T x_j$ for $1 \leq i,j \leq N_f$, which generalize the unique invariant of the bifundamental; these can be written as $\mathrm{Tr}(x B y^T)$ where $B$ is an $N_f \times N_f$ matrix representing a bilinear form; 
    \item The ``baryons" $\det (x_I)$ and $\det (y_I)$ for $I$ a subset of $\{1 , \dots , N_f\}$ of cardinality $N$.\footnote{We use the obvious notation where $x_I$ is the $N \times N$ matrix constructed from the $N$ vectors $x_i$ for $i \in I$. } 
\end{itemize}
In the case of $\widetilde{\mathrm{SU}}(N)$, this is modified as follows: 
\begin{itemize}
    \item The mesons transform under $\mathcal{P}$ as (\ref{mesonTransfoP}), which translates here in 
     \begin{equation}
     \label{mesonsSUtilde}
    \mathrm{Tr}(x B y^T) \rightarrow \mathrm{Tr}( A y B x^T (A^{-1})^T) = (-1)^{N-1} \mathrm{Tr}(x B^T y^T) \, .
\end{equation}
  In other words, the invariants are constructed from bilinear forms $B$ that satisfy 
  \begin{equation}
  \label{symmetryB}
      B=(-1)^{N-1} B^T \, , 
  \end{equation}
i.e. is symmetric for odd $N$ and antisymmetric for even $N$. 
    \item The baryons also transform under $\mathcal{P}$, as (recall (\ref{propertyA2})) 
    \begin{equation}
    \label{transformationBaryons}
        \det (x_I) \rightarrow \det (A y_I) = \det (y_I) \, , \qquad \det (y_I) \rightarrow \det (x_I) \, . 
    \end{equation}
    Therefore, the combination $\det (x_I) + \det (y_I) $ is invariant, and the combination $\det (x_I) - \det (y_I) $ is not. 
\end{itemize}
The full ring of invariants is quite intricate, as the example of $N=N_f=3$ shows, see equation (\ref{complicatedExample}) in the appendix. 

Finally, we turn to the global symmetry of this ring of invariants. When the gauge group is $\mathrm{SU}(N)$, it is well-known that we have a global symmetry $\mathrm{U}(N_f)$ acting as  
\begin{equation}
     \left( \begin{array}{c} x \\ y \end{array} \right) \rightarrow \left( \begin{array}{c} x U \\ y \bar{U} \end{array} \right)  \, . 
 \end{equation}
In the case of $\widetilde{\mathrm{SU}}(N)$, although the structure of the invariants is intricate, we can conjecture the global symmetry by inspection of the mesonic sector. Acting on the gauge invariant (\ref{mesonsSUtilde}), one obtains $ \mathrm{Tr}(x U B U^{\dagger} y^T) $ so that if $U B U^{\dagger} = B$ this is also a global invariant. Recalling (\ref{symmetryB}), we conjecture that the global symmetry group is (or at least, contains) $\mathrm{SO}(N_f)$ when $N$ is even, and $\mathrm{Sp}(\lfloor N_f/2\rfloor)$ when $N$ is odd. We will provide some consistency checks in section \ref{sectionHiggs}. We summarize these results in Table \ref{tableSummaryInvariantsNf}.

\begin{table}[t]
    \centering
    \begin{tabular}{|c|c|c|c|} \hline
       Invariants  & $\mathrm{SU}(N)$ & $\widetilde{\mathrm{SU}}(N)$, $N$ odd & $\widetilde{\mathrm{SU}}(N)$, $N$ even \\ \hline
       Baryons & $2 \binom{N_f}{N}$ & $\binom{N_f}{N}$ & $\binom{N_f}{N}$ \\   
        Mesons & $N_f^2$ & $\frac{1}{2}N_f(N_f+1)$ & $\frac{1}{2}N_f(N_f-1)$ \\  \hline
        Global symmetry & $\mathrm{U}(N_f)$ & $\mathrm{Sp}(\lfloor N_f/2\rfloor)$          & $\mathrm{SO}(N_f)$ \\  \hline
    \end{tabular}
    \caption{Number of mesons and baryons in the representation $N_f \otimes \mathrm{F \bar{F}}$ (with real dimension $2NN_f$), and associated global symmetry. }
    \label{tableSummaryInvariantsNf}
\end{table}


\subsubsection{Integration Formula}
\label{sectionIntegrationFormula}

We will need to be able to integrate over $\widetilde{\mathrm{SU}}(N)$, and we will use the Weyl integration formula. In this subsection, we explain in a very practical and concrete way how the integration is performed over $\widetilde{\mathrm{SU}}(N)$. For a more generic discussion, applicable to all principal extensions, we refer to \cite{wendt2001weyl}. 

Before going into this, let's introduce our Lie algebra conventions. To the basis of simple roots $\{\alpha_j\}$ (with $j=1 , \dots , N-1$, each having length squared $2$), we associate the basis of fundamental weights $\{ \varpi_j\}$ defined by $\langle  \alpha_i , \varpi_j \rangle = \delta_{ij}$. An integral weight is an element of $\mathfrak{h}^{\ast}$ of the form
\begin{equation}
    \lambda = \sum\limits_{i=1}^{N-1} \lambda_i \varpi_i \, ,
\end{equation}
where the $\lambda_i \in \mathbb{Z}$. To this weight, we associate the fugacity 
\begin{equation}
\label{defz}
   z(\lambda) = \begin{cases}
     \left( \prod\limits_{i=1}^{\frac{N-1}{2}} z_i^{\lambda_i + \lambda_{N-i}} \right)  \left(\prod\limits_{i=1}^{\frac{N-1}{2}} z_{\frac{N-1}{2}+i}^{\lambda_i - \lambda_{N-i}} \right)  & \textrm{ if } N \textrm{ is odd} \\ & \\
       \left( \prod\limits_{i=1}^{\frac{N}{2}-1} z_i^{\lambda_i + \lambda_{N-i}}  \right) \left( \prod\limits_{i=1}^{\frac{N}{2}-1} z_{\frac{N}{2}+i}^{\lambda_i - \lambda_{N-i}} \right) z_{\frac{N}{2}}^{\lambda_{\frac{N}{2}}} & \textrm{ if } N \textrm{ is even. }
    \end{cases}  
\end{equation}
This unconventional parametrization is chosen so that if $\mathcal{P}(\lambda) = \lambda$, then $z(\lambda)$ depends only on the $z_i$ for $i=1 , \dots , [N/2]$. This will play a crucial role in the following.

We introduce the measure for the maximal torus: 
\begin{equation}
\label{measuresu+}
     \d \mu_{N}^+ (z) = \prod\limits_{j=1}^{N-1} \frac{ \d z_j}{2 \pi i z_j} \prod\limits_{\alpha \in R^+(\mathfrak{su}(N))} \left( 1 - z(\alpha)\right) \, ,
\end{equation}
where $R^+(\mathfrak{su}(N))$ is the set of the $\frac{1}{2}N(N-1)$ positive roots of $\mathfrak{su}(N)$. The superscript $+$ in $\d \mu_{N}^+ (z)$ is introduced here for later convenience. The Weyl integration formula over $\mathrm{SU}(N)$ then states that for a class function $f : \mathrm{SU}(N) \rightarrow \mathbb{C}$, we have 
\begin{equation}
\label{IntegrationSUN}
    \int_{\mathrm{SU}(N)} \d \eta_{\mathrm{SU}(N)}(X)f(X) =  \oint_{|z_j| = 1}  \d \mu_{N}^+ (z)f(z)   \,  . 
\end{equation}
Here and in the following formulas, $\d \eta_{G}$ is the Haar measure for the group $G$.

Now we review how this formula should be modified to yield integration over $\widetilde{\mathrm{SU}}(N)$. For that, need to take into account the other connected component of the group. It turns out the elements of this components can also be diagonalized, but the measure on the maximal torus is no longer (\ref{measuresu+}). Rather, it is another measure $\d \mu_{N}^- (z)$ that we now describe. We have to study separately both parities of $N$. 
\begin{itemize}
\item $N$ even. There is an odd number of simple roots, and the $\mathcal{P}$-invariant subspace of $\mathfrak{h}^{\ast}$ is spanned by 
\begin{equation}
    \beta_j = \alpha_j + \alpha_{N-j} \, \qquad \left(j=1 , \dots , \frac{N}{2}-1\right) \, , \qquad \beta_{N/2} = \alpha_{N/2} \, . 
\end{equation}
One can check that the $\{\beta_j\}$, $j=1 , \dots , \frac{N}{2}$ are the simple roots of root system of type $B_{N/2}$ (the norm squared of the roots is 2 for the short roots and 4 for the long roots). Let $R^+(B_{N/2})$ be the set of positive roots in this system. We introduce the measure 
\begin{equation*}
     \d \mu_{N}^- (z)= \prod\limits_{j=1}^{N-1} \frac{ \d z_j}{2 \pi i z_j} \prod\limits_{\alpha \in R^+(B_{N/2})} \left( 1 - z(\alpha)\right) \, . 
\end{equation*}
By construction, the product does not depend on the $z_i$ for $i > N/2$, so we will identify this measure with the one involving only the $z_i$ for $i \leq N/2$, and write abusively 
\begin{equation}
\label{measuresu-even}
    N \textrm{ even:} \qquad  \d \mu_{N}^-(z) = \prod\limits_{j=1}^{N/2} \frac{ \d z_j}{2 \pi i z_j} \prod\limits_{\alpha \in R^+(B_{N/2})} \left( 1 - z(\alpha)\right) \, . 
\end{equation}
    \item $N$ odd. In this case, the construction is slightly more involved. The set of roots of $\mathfrak{su}(N)$ projected to the $\mathcal{P}$-invariant subspace of $\mathfrak{h}^{\ast}$ contains elements of three distinct length squared, $1/2$, $1$ and $2$.\footnote{This is strictly true only for $N \geq 5$. For instance $\frac{1}{2}(\alpha_{(N-1)/2} + \alpha_{(N+1)/2})$ has length squared $1/2$, $\frac{1}{2}(\alpha_{(N-2)/2} + \alpha_{(N+2)/2})$ has length squared $1$ and $\alpha_{(N-1)/2} + \alpha_{(N+1)/2}$ has length squared $2$. The case $N=3$ has no element of length squared $1$; it will be treated in section \ref{sectionsu3}.  } One can show that the \emph{doubles} of all the elements of this set with length squared $1$ and $2$ form a root system of type $C_{(N-1)/2}$, in which the long roots have length squared $8$ and the short roots have length squared $4$. We call $R(C_{(N-1)/2})$ this root system, and define    
    \begin{equation}
    \label{measuresu-odd}
    N \textrm{ odd:} \qquad  \d \mu_{N}^- (z)= \prod\limits_{j=1}^{(N-1)/2} \frac{ \d z_j}{2 \pi i z_j} \prod\limits_{\alpha \in R^+(C_{(N-1)/2})} \left( 1 - z(\alpha)\right) \, . 
\end{equation}
\end{itemize}

We are now ready to state the Weyl integration formula for $\widetilde{\mathrm{SU}}(N)$. Consider a function $f$ invariant under conjugation, which means $f(YXY^{-1}) = f(X)$ for all $X,Y \in \widetilde{\mathrm{SU}}(N)$.
Then \cite{wendt2001weyl}
\begin{equation}
\label{eq:integration}
    \int_{\widetilde{\mathrm{SU}}(N)} \d \eta_{\widetilde{\mathrm{SU}}(N)}(X)f(X) = \frac{1}{2} \left[ \oint_{|z_j|=1}  \d \mu_{N}^+ (z)f(z) + \oint_{|z_j|=1}  \d \mu_{N}^- (z) f( z^{\mathcal{P}}) \right] \,  . 
\end{equation}

For reference, we give the values of the $z(\alpha)$ that appear in the various formulas above, for small values of $N$, in Table \ref{tablezalpha}. 

\begin{table}[t]
    \centering
    \begin{tabular}{|c|c|}
    \hline 
        $N$ & Values of $z(\alpha)$ \\  \hline 
        $2$ & \begin{tabular}{c}
            $z_1^2$  \\ $z_1^2$
        \end{tabular} \\   \hline 
        $3$ & \begin{tabular}{c}
            $z_ 1 z_ 2^3,z_ 1^2,\frac{z_ 1}{z_ 2^3}$  \\ $z_1^4$
        \end{tabular} \\   \hline 
        $4$ & \begin{tabular}{c}
           $\frac{z_1^2 z_3^2}{z_2},z_2   z_3^2,\frac{z_2^2}{z_1^2},z_1^2,\frac{z_2}{z   _3^2},\frac{z_1^2}{z_2 z_3^2}$  \\ $\frac{z_1^4}{z_2^2},z_2^2,\frac{z_2^2}{z   _1^2},z_1^2$ 
        \end{tabular} \\   \hline 
          $5$ & \begin{tabular}{c}
           $\frac{z_1^2 z_3^2}{z_2 z_4},z_1 z_3   z_4^2,\frac{z_2 z_4^3}{z_1 z_3},\frac{z_2   z_3^2}{z_4},\frac{z_2^2}{z_1^2},\frac{z_2   z_3}{z_1 z_4^3},z_1^2,\frac{z_2   z_4}{z_3^2},\frac{z_1}{z_3   z_4^2},\frac{z_1^2 z_4}{z_2 z_3^2}$  \\ $\frac{z_1^4}{z_2^2},z_2^2,\frac{z_2^4}{z   _1^4},z_1^4$ 
        \end{tabular} \\   \hline 
             $6$ & \begin{tabular}{c}
           $\frac{z_1^2 z_4^2}{z_2 z_5},\frac{z_1   z_2 z_4 z_5}{z_3},\frac{z_2^2 z_5^2}{z_1 z_3   z_4},\frac{z_1 z_3 z_4 z_5}{z_2},\frac{z_3   z_5^2}{z_1   z_4},\frac{z_3^2}{z_2^2},\frac{z_2   z_4^2}{z_5},\frac{z_2^2}{z_1^2},\frac{z_3   z_4}{z_1 z_5^2},\frac{z_2^2 z_4}{z_1 z_3   z_5^2},z_1^2,\frac{z_2 z_5}{z_4^2},\frac{z_1   z_3}{z_2 z_4 z_5},\frac{z_1 z_2}{z_3 z_4   z_5},\frac{z_1^2 z_5}{z_2 z_4^2}$  \\ $\frac{z_1^4}{z_2^2},\frac{z_1^2   z_2^2}{z_3^2},\frac{z_2^4}{z_1^2   z_3^2},\frac{z_1^2   z_3^2}{z_2^2},\frac{z_3^2}{z_1^2},\frac{z_3^   2}{z_2^2},z_2^2,\frac{z_2^2}{z_1^2},z_1^2$ 
        \end{tabular} \\   \hline 
    \end{tabular}
    \caption{This table gives the values of $z(\alpha)$ for the positive roots $\alpha$ appearing in the measures, for small values of $N$. In each case, the first line corresponds to the roots in equation (\ref{measuresu+}), and the second line corresponds to the roots of (\ref{measuresu-even}) and (\ref{measuresu-odd}), depending on the parity of $N$. When $N=2$, the two measures are identical, reflecting the fact that $\widetilde{\mathrm{SU}}(2) = \mathrm{SU}(2)$. }
    \label{tablezalpha}
\end{table}

\subsubsection{\texorpdfstring{The $\widetilde{\mathrm{SU}}(3)$ example}{The \mathrm{SU}(3) tilde example}}
\label{sectionsu3}

In this subsection, we illustrate all the abstract constructions of the previous paragraphs in the simplest non-trivial case, $N=3$. 

Let us begin with the representations of $\mathrm{SU}(3)$ With the somewhat unconventional choice (\ref{defz}), the character of the fundamental, antifundamental and adjoint representations of $\mathrm{SU}(3)$ are
\begin{equation}
    \chi_{\textrm{F}} = z_1 z_2 + \frac{1}{z_2^2} + \frac{z_2}{z_1}  \, , 
\end{equation}
\begin{equation}
    \chi_{\bar{\textrm{F}}} =\frac{z_1}{z_2} + \frac{1}{z_1 z_2} + z_2^2  \, , 
\end{equation}
\begin{equation}
\label{adjSU3}
    \chi_{\textrm{Adj}} = 2+ z_1^2 + \frac{z_1}{z_2^3} + z_1 z_2^3 + \frac{1}{z_1^2} + \frac{z_2^3 }{z_1} + \frac{1}{z_1 z_2^3}\, . 
\end{equation}
As for $\widetilde{\mathrm{SU}}(3)$ representations, the adjoint character is still given by (\ref{adjSU3}) while the bifundamental is 
\begin{equation}
    \chi_{\textrm{F}\bar{\textrm{F}}} =  \chi_{\textrm{F}} +  \chi_{\bar{\textrm{F}}} \, . 
\end{equation}

For the integration measures, we read in Table \ref{tablezalpha} that  
\begin{equation}
\label{su3measure}
    \d \mu_{3}^+ (z)= \frac{\d z_1}{2 \pi i z_1 } \frac{\d z_2}{2 \pi i z_2 } \left( 1- z_1^2 \right)  \left( 1-  \frac{z_1}{z_2^3} \right)  \left( 1- z_1 z_2^3 \right)  \, . 
\end{equation}
\begin{equation}
\label{2C1measure}
    \d \mu_{3}^- (z)= \frac{\d z_1}{2 \pi i z_1 } \left( 1- z_1^4 \right) \, . 
\end{equation}

Now let us describe the action of $\mathcal{P}$ concretely. If we choose\footnote{We use the standard notation $E_{\alpha}$ for a generator of the root space associated with the root $\alpha$, normalized bu the Killing form. } 
\begin{equation}
    (H_1 , H_2 , E_{\alpha_1} , E_{-\alpha_1} ,E_{\alpha_2} , E_{-\alpha_2} , E_{\alpha_1 + \alpha_2} ,E_{-\alpha_1 - \alpha_2} )
\end{equation}
for the basis of $\mathfrak{su}(3)$, in this order, the diagonal matrix $\Phi_{\textrm{Adj}} (z)$ follows directly from the character of the adjoint representation (\ref{adjSU3}), while $ \Phi_{\textrm{Adj}} (\mathcal{P})$ is a non-diagonal matrix which implements the action of $\mathcal{P}$ in the same basis:
\begin{eqnarray}
\label{PhiAdjsu3}
    \Phi_{\textrm{Adj}} (z) &=& \mathrm{Diag} \left(1,1, z_1 z_2^3,\frac{1}{z_1 z_2^3}, \frac{z_1}{z_2^3} , \frac{z_2^3 }{z_1} , z_1^2  ,\frac{1}{z_1^2}  \right) \, ,  \nonumber \\
    \Phi_{\textrm{Adj}} (\mathcal{P}) &=& \left(
\begin{array}{cccccccc}
 0 & 1 & 0 & 0 & 0 & 0 & 0 & 0 \\
 1 & 0 & 0 & 0 & 0 & 0 & 0 & 0 \\
 0 & 0 & 0 & 0 & 1 & 0 & 0 & 0 \\
 0 & 0 & 0 & 0 & 0 & 1 & 0 & 0 \\
 0 & 0 & 1 & 0 & 0 & 0 & 0 & 0 \\
 0 & 0 & 0 & 1 & 0 & 0 & 0 & 0 \\
 0 & 0 & 0 & 0 & 0 & 0 & -1 & 0 \\
 0 & 0 & 0 & 0 & 0 & 0 & 0 & -1 \\
\end{array}
\right)
    \, . 
\end{eqnarray}
Note that the minus signs in the above matrix are consequences of the $(-1)^{1+i-j}$ in (\ref{actionPCcomplexifiedAlg}).

\subsection{\texorpdfstring{The group $\widetilde{\mathrm{SO}}(2N)$}{The group SO(2N) tilde}}

As we explained in \cite{Bourget:2017tmt,Bourget:2017sxr}, the principal extension has a natural incarnation when dealing with algebras of type $D_N$. Indeed, as mentioned above, in this case the principal extension is the disconnected group $\mathrm{O}(2N)$. The integration formula \cite{wendt2001weyl} then takes the form 
\begin{equation}
\label{OnIntegration}
     \int_{\mathrm{O}(2n)} \d \eta_{\mathrm{O}(2n)}(X) f(X) = \frac{1}{2} \left[  \int \d \eta_{\mathrm{SO}(2n)} (z) f(z) + \int  \d \eta_{\mathrm{Sp}(n-1)} (z) f(z^{\mathcal{P}} )  \right]  \, . 
\end{equation}
We refer to \cite{Bourget:2017tmt,Bourget:2017sxr} for the details, in particular concerning the integration measures.

\section{\texorpdfstring{Supersymmetric gauge theory with gauge group $\widetilde{G}$}{Supersymmetric gauge theory with gauge group G tilde}}\label{SUSYtheories}

Having introduced the family of principal extensions on purely mathematical grounds, it is then natural to wonder about the potential use of these groups in physical theories. In particular, one may ask whether it is possible to use these groups as the starting point to construct new gauge theories. To begin with, note that principal extensions $\widetilde{G}$ are, in the end, nothing but Lie groups. Thus we can apply the textbook construction of gauge theories based on $\widetilde{G}$. Hence, in particular, the gauge theories to be constructed are nothing but standard Lagrangian theories. Moreover, the construction (at the classical level) is independent of the space-time dimension, and hence we may consider generic gauge theories in any dimension. For the sake of concreteness, in this work we will restrict our attention to four-dimensional $\mathcal{N}=2$ theories, and we will concentrate on gauge groups which are principal extensions of $G=\mathrm{SU}(N)$. 

As described, $\widetilde{G}$ can be thought simply as particular examples of Lie groups. Hence the construction of the corresponding $\mathcal{N}=2$ gauge theory follows from the standard techniques (for instance, see \cite{Tachikawa:2013kta} for a textbook approach). Thus, the basic ingredients will be $\mathcal{N}=2$ vector multiplets in the adjoint of $\widetilde{G}$ (representation introduced in the previous section) and hypermultiplets in some representation $\mathcal{R}$ of $\widetilde{G}$, with the standard $\mathcal{N}=2$ coupling between those. In the remainder of the paper we will consider the particular case of SQCD-like theories, where $\mathcal{R}$ is the fundamental representation of $\widetilde{G}$ (representation introduced as well in the previous section), and we will refere to these theories as the $\widetilde{G}$ theories (as opposed to the same theory based on the $G$ group, to which we will refere as the $G$ theory).

Note that the associated Lie algebra to $\widetilde{G}$ is just the same as that of $G$. Hence the perturbative quantization of the $\widetilde{G}$ theory --which only sees small field configurations close to the origin in field space, and thus is only sensitive to the connected component of the principal extension-- is identical to that of the $G$ theory. Therefore the local dynamics of the theories based on gauge groups $G$ and $\widetilde{G}$ are the same (as emphasized recentely in \textit{e.g.} \cite{Aharony:2016kai,Argyres:2016yzz}), and in particular, the Feynman rules are identical in both theories, and hence the computation of the $\beta$ function is just analogous as in the case of the $G$ theory.\footnote{Indeed, to find the Feynman rules --on $\mathbb{R}^4$-- one would expand the lagrangian perturbatively about the vacuum. The Feynman rules are then read off from correlators in the free theory, which by construction is only sensitive to the connected component of the group. Hence, the effect of the disconnected component is to impose ``superselection rules" on the available operators, yet the dynamics remains unchanged.} An important consequence of this is that the conditions for conformality must be just the same as in the $G$ case. In the particular case of SQCD-like theories, it then follows that superconformal invariance requires the $\widetilde{\mathrm{SU}}(N)$ to have $2N$ $\mathrm{SU}(N)$ fundamental hypermultiplets.

Another consequence of falling into the standard Lagrangian class is that $\widetilde{G}$ theories are free of local anomalies.\footnote{To begin with, local anomalies are directly absent in $\mathcal{N}=2$ theories due to the fact that SUSY requires the use of real representations. In the case of $\widetilde{SU}(N)$ theories the representations we are using are, to begin with, real, and hence the very same argument applies --note that in fact it applies more widely, even to less SUSY cases.} Indeed, since local dynamics is unchanged under discrete gauging, the values of the central charges $a$ and $c$ are identical in the $G$ and $\widetilde{G}$ theories -- this property was already used in \cite{Argyres:2016yzz}. Nevertheless, it would be very interesting to further elaborate on this point, studying in particular the global aspects of $\widetilde{G}$ and whether global anomalies may arise.

As described in the previous section, the $\widetilde{G}$ group is a disconnected Lie group roughly speaking obtained by enlarging the gauge group by automorphisms of the Dynkin diagram. In the case of $\widetilde{\mathrm{SU}}(N)$ these amount to the invariance under flipping of the Dynkin diagram --operation denoted in the previous section by $\mathcal{P}$, which, together with the identity, forms the automorphism group $\Gamma$ of the Dynkin diagram. Note that, in particular, $\mathcal{P}$ exchanges the fundamental and antifundamental representations of the underlying $\mathrm{SU}(N)$, and hence it is essentially like the charge conjugation symmetry $\mathcal{C}$. Thus the $\widetilde{G}$ theory can be thought as an implementation of gauging charge conjugation symmetry. Note that $\widetilde{G}$ is not simply the direct product of the connected group times the discrete group $\Gamma$, i.e. $\widetilde{G}\ne G\times \Gamma$; but rather the semidirect product of these, i.e. $\widetilde{G}= G\rtimes \Gamma$. There has been recent attempts \cite{Argyres:2016yzz} to produce theories by further gauging the charge conjugation symmetry in a standard $\mathcal{N}=2$ gauge theory\footnote{Gauging discrete symmetries, leading to disconnected gauge groups, plays also a very important role in constructing $\mathcal{N}=3$ SCFT's. See e.g. \cite{Garcia-Etxebarria:2017ffg,Garcia-Etxebarria:2015wns,Aharony:2016kai,Lemos:2016xke,Argyres:2018wxu,Tom}.}, a construction analogous to having considered an enlarged gauge group $G\times \Gamma$. These works mostly focus on rank one theories, finding that generically it is quite hard to find $\mathcal{N}=2$ consistent such gaugings. Instead, our approach allows for a version of such gauging which is automatically consistent for arbitrary rank, since, at the end of the day, the semidirect product structure ensures that the resulting structure $\widetilde{G}$ is just a (non-connected) Lie group, in such a way that the standard construction of Lagrangian $\mathcal{N}=2$ theories goes through essentially unchanged. Note that the case of $\widetilde{\mathrm{SU}}(2)$ is trivial, and $\widetilde{\mathrm{SU}}(2)\equiv \mathrm{SU}(2)$. Hence our results are compatible with the existing classifications of rank one theories (see e.g. \cite{Argyres:2016xua,Argyres:2016yzz}), as the $\widetilde{\mathrm{SU}}(2)$ is just the same as the familiar $\mathrm{SU}(2)$ $\mathcal{N}=2$ gauge theory with 4 flavors.

Since, at the end of the day, our $\widetilde{G}$ theories are particular examples of (conformal) $\mathcal{N}=2$ theories, we may use the powerful techniques developed over the last few years to study exact aspects of them. In particular we will make extensive use of the $\mathcal{N}=2$ superconformal index (SCI). In the conventions of \cite{Gadde:2011uv}, the index for a Lagrangian theory based on some Lie group $G^{\circ}$ -- which, for us, will be either $G=\mathrm{SU}(N)$ or $\widetilde{G}=\widetilde{\mathrm{SU}}(N)$ -- reads (let us unrefine the global symmetry to ease the presentation)
\begin{equation}
\label{SCI}
    \mathcal{I}_{G^{\circ}}=\int  \d \eta_{G^{\circ}}(X)\,{\rm PE}\Big[\sum_{i\in {\rm multiplets}} f^{\mathcal{R}_i}\,\chi_{\mathcal{R}_i}(X)\Big]\, ;
\end{equation}
where the sum inside the ${\rm PE}$ (which stands for the plethystic exponential\footnote{The plethystic exponential of a function $f$, such that $f(0)=0$, is defined as ${\rm PE}[f(\cdot)]=e^{\sum_{n=1}^{\infty}\frac{f(\cdot^n)}{n}}$.}) runs to all multiplets in the theory, each in a representation $\mathcal{R}_i$ of the gauge group whose character is $\chi_{\mathcal{R}_i}$. The precise contributions of the vector multiplets and the (half) hypermultiplets are 
\begin{equation}
\label{eq:singleletter}
    f^V=-\frac{\sigma\,\tau}{1-\sigma\,\tau}-\frac{\rho\,\tau}{1-\rho\,\tau}+\frac{\sigma\,\rho-\tau^2}{(1-\rho\,\tau)\,(1-\sigma\,\tau)}\,, \qquad 
    f^{\frac{1}{2}H}=\frac{\tau\,(1-\rho\,\sigma)}{(1-\rho\,\tau)\,(1-\sigma\,\tau)}\, .
\end{equation}
Finally, $\d\eta_{G^{\circ}}(X)$ stands for the appropriate Haar measure on the corresponding gauge group. Both the measure and the implementation of the characters for the $\widetilde{G}$ group have been explicitly described above, while those for the $G$ theory are the standard ones. Hence the evaluation of the SCI for either the $G$ or $\widetilde{G}$ theory is in principle a straightforward task. Nevertheless, the SCI is a complicated function of all fugacities. In order to extract more direct information, we will focus on a particular limit with enhanced SUSY \cite{Gadde:2011uv}, namely the Coulomb branch index, whereby we take $\tau\rightarrow 0$ with $\sigma$ and $\rho$ fixed. In this limit, only the vector multiplet contributes, and on general grounds this is expected to be a counting of the Coulomb branch invariants of the group.

The second tool which we will use is the Higgs branch Hilbert series \cite{Benvenuti:2006qr,Benvenuti:2010pq}. Note that standard arguments of $\mathcal{N}=2$ SUSY ensure the Higgs branch to be absent of quantum corrections. Hence, for the analysis of the Higgs branch we may drop conformality and actually consider a wider set of theories with gauge group $\widetilde{\mathrm{SU}}(N)$ and $N_f$ flavors. Then, the Higgs branch Hilbert series schematically reads
\begin{equation}
    HS_{(N,\,N_f)}=\int \d \eta_G\,{\rm PE}\Big[\sum_{i\in {\rm Higgs}} t^{\Delta_i}\,\chi_{\mathcal{R}_i}-\sum_{F^{\#}_{\rm Adj}}t^2\,\chi_{\rm Adj}\Big]\,;
\end{equation}
where the first sum inside the plethystic exponential runs to all fields in matter multiplets in representations $\mathcal{R}_i$ while the second sum runs to all F-term constraints to impose so as to be in the Higgs branch -- i.e. the F-terms of adjoint scalars in vector multiplets. Moreover, both the integration measure as well as the characters of the gauge group to be used in the evaluation of the Higgs branch Hilbert series have been, just like their SCI counterparts, described explicitly above.

The Hilbert series is a counting of chiral operators in the Higgs branch of the theory at the superconformal fixed point, and as such it is sensitive to the global flavor symmetry group of the gauge theory. This will make it a very useful tool to study the global symmetries in $\widetilde{G}$ theories. It should be noted that, at least in the particular case of class $\mathcal{S}$ theories of genus zero, the Higgs branch Hilbert series coincides \cite{Gadde:2011uv,Gaiotto:2012uq} with the Hall-Littlewood limit of the index, which is obtained by taking $\sigma,\,\rho\rightarrow 0$ while keeping $\tau$ fixed. While it is unknown whether $\widetilde{G}$ theories fit into class $\mathcal{S}$, the very same argument which allows to embed the Higgs branch Hilbert series as a Hall-Littlewood index still applies, since formally and for the theories at hand the vector multiplet contribution to the Hall-Littlewood index is identical to the constraint contribution to the Hilbert series (the numerator) while the matter contribution to the Hall-Littlewood index is identical to the matter contribution to the Hilbert series (the numerator).

\section{The non-freely generated Coulomb branch}\label{CB}

\label{sec:Coulomb}
As described above, the Coulomb limit of the superconformal index is obtained taking the limit \cite{Gadde:2011uv}
\begin{equation*}
    \tau \rightarrow 0,  \ \ \rho,\sigma \ \ \textrm{fixed} \, .
\end{equation*}
Therefore the single letter partition functions (\ref{eq:singleletter}) reduce to $f^{\frac{1}{2}H} = 0$ and $f^{V} = \rho\sigma =: t$. Thus, only the vector multiplet contributes to this limit of the index, which then becomes a counting of the invariants in the Coulomb branch. In fact, this equivalent to the computation of the Hilbert series of the Coulomb branch, and can be recast as
\begin{equation}
\label{eq:cindex}
    \mathcal{I}^{\textrm{Coulomb}}_{G^{\circ}}(t) = \int_{G^{\circ}} \d \eta_{G^{\circ}} (X) \frac{1}{\mathrm{det} \left( 1 - t \Phi_{\mathrm{Adj}} (X)\right)} \, ,
\end{equation}
where $\d \eta_{G^{\circ}}(X)$ is the Haar measure of the gauge group, while $\Phi_{\mathrm{Adj}}$ denote the adjoint representation (i.e. ${\rm Tr}(\Phi_{\mathrm{Adj}})$ would be the character of the adjoint representation). 

Upon evaluating (\ref{eq:cindex}) for theories of type $A_n$ with gauge group $\widetilde{\mathrm{SU}}(N)$ for the first few $N$'s, one can convince oneself that 
\begin{equation}
\label{ICoulomb}
 \mathcal{I}^{\textrm{Coulomb}}_{\widetilde{\mathrm{SU}}(N)}(t) = \frac{1}{2}\left[\prod\limits_{i=2}^{N}\frac{1}{1-t^i}+\prod\limits_{i=2}^{N}\frac{1}{1-(-t)^i}\right] \, .
 \end{equation}
 One can rewrite this in the form 
 \begin{equation}
     \mathcal{I}^{\textrm{Coulomb}}_{\widetilde{\mathrm{SU}}(N)}(t) = \frac{ \sum\limits_{k_1 < \dots < k_r \, \mathrm{odd}} t^{k_1 + \dots + k_r}}{\prod\limits_{i \, \mathrm{even}} (1-t^i) \prod\limits_{i \, \mathrm{odd}} (1-t^{2i})}  \, ,
 \end{equation}
 where all the indices are between $2$ and $N$ and $r \geq 0$ is even. This makes manifest the structure of primary (in the denominator) and secondary (in the numerator) invariants described in section \ref{sectionRepandInv}. As soon as the numerator is non-trivial (i.e different from $1$), the Coulomb branch is not freely generated --as observed in section \ref{sectionGtilde}, this is the case for $N \geq 5$--. This is on very general grounds, since the numerator stands for relations while the denominator stands for generators of the ring of invariants (see  \cite{Benvenuti:2006qr} for a detailed explanation).
 
 This can be observed explicitly on the plethystic logarithm (see appendix \ref{AppendixInvariant}) of (\ref{ICoulomb}), which we reproduce here for the first values of $N$:
 \begin{center}
     \begin{tabular}{|c|c|}  \hline
     $N$ & PL of $ \mathcal{I}^{\textrm{Coulomb}}_{\widetilde{\mathrm{SU}}(N)}(t)$ \\ \hline
 2 & $t^2$ \\
 3 & $t^2+t^6$ \\
 4 & $t^2+t^4+t^6$ \\
 5 & $t^2+t^4+t^6+t^8+t^{10}-t^{16}$ \\
 6 & $t^2+t^4+2 t^6+t^8+t^{10}-t^{16}$ \\
 7 & $t^2+t^4+2 t^6+t^8+2 t^{10}+t^{12}+t^{14}-t^{16}-t^{18} + \dots$ (infinite) \\  \hline
     \end{tabular}
 \end{center}
For $N < 5$, the plethystic logarithm is a polynomial with positive integer coefficients, corresponding to free generators. Then starting at $N=5$ the Coulomb branch index starts showing relations among generators \cite{Benvenuti:2006qr} accounted for by the terms with negative coefficients (note that those, when put in the ${\rm PE}$, contribute to the numerator, thus supporting our claim above. In turn, positive contributions to the ${\rm PL}$ corresponding to generators contribute to the denominator once the ${\rm PE}$ is taken). Thus we come to the surprising conclusion that the Coulomb branch of the $\widetilde{G}$ theories is not freely generated. In fact, for $N\geq 7$ not only such negative terms arise but also the plethystic logarithm becomes a meromorphic function (hinted in the fact that the plethystic logarithm has an infinite series expansion, instead of being just a polynomial), showing also that the Coulomb branch is not only non-freely generated but also a non-complete intersection.

While not ruled out by any argument, the lack of explicit examples of theories with non-freely generated Coulomb branches has created the standard lore that no such theory exist (see e.g. \cite{Argyres:2017tmj} for a recent discussion). The $\widetilde{G}$ theories provide counterexamples to such standard lore by explicitly realizing $\mathcal{N}=2$ theories with a non-freely generated Coulomb branch.\footnote{Note that the results in \cite{Argyres:2017tmj} do not directly apply to our examples, since that reference concerns rank one theories, while our examples of $\widetilde{G}$ non-freely generated would start at rank 5.} In retrospect one can check in the mathematical literature that whenever one considers disconnected groups, very frequently the ring of invariants stops being freely generated. In fact, using \cite{chevalley1955invariants,Schmelkin}, one can see that the principal extension groups $\widetilde{\mathrm{SU}}(N)$ must indeed produce non freely generated rings in the adjoint for $N>4$. Indeed, the criterion for the ring to be freely generated is that the component group $\Gamma$ should act by reflections (a symmetry with codimension one kernel) on the invariant ring of the connected components. Here the action (\ref{AdjointP}) has codimension $\leq 1$ kernel if and only if $N \leq 4$. 

Note that when applying this very same logic to the principal extension of $\mathrm{SO}(2N)$ groups -- \textit{a.k.a.} $\mathrm{O}(2N)$ --, the ring of invariants is, in that case, freely generated. This nicely matches with the fact that $\widetilde{\mathrm{SU}}(4)=\mathrm{O}(6)$, so that internal consistency demands that the case $N=4$ must be freely generated -- and indeed this is the largest $N$ for which it is. 

The Coulomb branch limit of the SCI is, in a sense, a very rough observable, insensitive, for instance, to the global symmetry of the theory. Thus, in order to elucidate on that, in the next section we turn to the Higgs branch and compute the Higgs branch Hilbert series -- which can be regarded, in our case, as the Hall-Littlewood limit of the SCI --, sensitive to the global symmetry. Note that, being such a rough observable, the Coulomb branch index of SQCD would coincide with the Coulomb branch index of, for instance, the $\mathcal{N}=4$ SYM theory based on the same $\widetilde{SU}(N)$. Note as well that our Coulomb branch index coincides formally with that found in \cite{Argyres:2018wxu,Tom}, yet the latter for a completely different theory in an unrelated context. One may speculate that the reason for such ``accidental" agreement (because, as emphasized, the theories are completely different) is that the Coulomb branch index is a rough observable which is only sensitive to the Coulomb branch structure, so from its point of view an $\mathcal{N}=3$ theory with the same gauged discrete symmetries is indistinguishable from our $\mathcal{N}=2$ theories (in the same sense that \textit{e.g.} $SU(N)$ $\mathcal{N}=4$ SYM has the same Coulomb branch index as $\mathcal{N}=2$ $SU(N)$ superconformal QCD).

\section{Higgs branch and global symmetry of SQCD}
\label{sectionHiggs}

Let us now turn to the Higgs branch of the $\widetilde{G}$ theories, where the global non-R symmetry must be manifest. As explained in section \ref{SUSYtheories} to study it we use the Higgs branch Hilbert series \cite{Benvenuti:2006qr,Benvenuti:2010pq}. This can be seen as another limit of the index (\ref{SCI}) with $\sigma , \rho \rightarrow 0$ and $t:= \tau$ fixed \cite{Gadde:2011uv}. In the following we compute this for the $\widetilde{\mathrm{SU}}(N)$ theories with $N_f$ fundamental hypermultiplets, restricting to the case $N_f \geq 2 N$. Then, the general form of the Higgs branch Hilbert series can be recast as (see also \cite{derksen2015computational})
\begin{equation}
\label{HiggsBranchFormula}
   HS_{(N,\,N_f)} = \int_{G^{\circ}} \d \eta_{G^{\circ}} (X) \frac{\mathrm{det} \left( 1 - t^2 \Phi_{\mathrm{Adj}} (X)\right)}{\mathrm{det} \left( 1 - t \Phi_{\mathrm{F\bar{F}}} (X)\right)^{N_f} } \,  ,
\end{equation}
where $d\eta_{G^{\circ}}(X)$ is the gauge group Haar measure, while $\Phi_{\mathrm{Adj}}$ and $\Phi_{\mathrm{F\bar{F}}}$ denote the adjoint representation and the bifundamental representation respectively. The denominator of the integrand accounts for the chiral multiplets, which are the fields whose vacuum expectation value contribute to the Higgs branch, and the numerator stands for the F-terms, which transform in the adjoint representation of the gauge group. Then the integration projects to gauge invariants. The formula (\ref{HiggsBranchFormula}) is valid under the assumption that the F-terms define a complete intersection, which is the case when $N_f \geq 2 N$, hence our restriction to this window. 

Let us now put this into practise by computing the first orders of the $t$ expansion for the lowest values of $(N,\,N_f)$, see Table \ref{TableHSQCD}. For completeness, we also display the result for the connected component alone (denoted by $HS^+_{(N,\,N_f)}$), which is equal to the Higgs branch Hilbert series of SQCD with $\mathrm{SU}(N)$ gauge group and $N_f$ flavors. As a simple consistency check, one observes the general pattern of Table \ref{tableSummaryInvariantsNf} in the coefficient of $t^2$, which equals $N_f^2$ when the gauge group is $\mathrm{SU}(N)$ and $\frac{1}{2}N_f (N_f + (-1)^{N-1})$ when it is $\widetilde{\mathrm{SU}}(N)$. 

\begin{table}[t]
    \centering
\begin{equation*}
    \begin{array}{|c|c|} \hline
     (N,N_f) & 
\begin{array}{c}
 HS^{+}_{(N,N_f)} \\
  HS_{(N,N_f)}  \\ 
\end{array}
 \\ \hline
 (3,6) & 
\begin{array}{c}
 1+36 t^2+40 t^3+630 t^4+1120 t^5+7525 t^6+O\left(t^7\right) \\ 
 1+21 t^2+20 t^3+336 t^4+560 t^5+3850 t^6+O\left(t^7\right) \\
\end{array}
\\ \hline
 (3,7) &  
\begin{array}{c}
 1+49 t^2+70 t^3+1176 t^4+2716 t^5+19452 t^6+O\left(t^7\right) \\ 
 1+28 t^2+35 t^3+616 t^4+1358 t^5+9862 t^6+O\left(t^7\right) \\
\end{array}
 \\ \hline
 (3,8) & 
\begin{array}{c}
 1+64 t^2+112 t^3+2016 t^4+5712 t^5+44079 t^6+O\left(t^7\right) \\ 
 1+36 t^2+56 t^3+1044 t^4+2856 t^5+22239 t^6+O\left(t^7\right) \\
\end{array}
\\ \hline
 (3,9) & 
\begin{array}{c}
 1+81 t^2+168 t^3+3240 t^4+10860 t^5+90440 t^6+O\left(t^7\right) \\
 1+45 t^2+84 t^3+1665 t^4+5430 t^5+45500 t^6+O\left(t^7\right) \\
\end{array}
 \\ \hline
 (3,10) & 
\begin{array}{c}
 1+100 t^2+240 t^3+4950 t^4+19140 t^5+171699 t^6+O\left(t^7\right) \\
 1+55 t^2+120 t^3+2530 t^4+9570 t^5+86229 t^6+O\left(t^7\right) \\
\end{array}
 \\ \hline
 (4,8) & 
\begin{array}{c}
 1+64 t^2+2156 t^4+49035 t^6+O\left(t^7\right) \\
 1+28 t^2+1106 t^4+24381 t^6+O\left(t^7\right) \\
\end{array}
 \\ \hline
 (4,9) & 
\begin{array}{c}
 1+81 t^2+3492 t^4+102284 t^6+O\left(t^7\right) \\
 1+36 t^2+1782 t^4+50942 t^6+O\left(t^7\right) \\
\end{array}
 \\ \hline
 (4,10) & 
\begin{array}{c}
 1+100 t^2+5370 t^4+196779 t^6+O\left(t^7\right) \\
 1+45 t^2+2730 t^4+98109 t^6+O\left(t^7\right) \\
\end{array}
 \\ \hline
 (5,10) & 
\begin{array}{c}
 1+100 t^2+4950 t^4+504 t^5+161799 t^6+O\left(t^7\right) \\
 1+55 t^2+2530 t^4+252 t^5+81279 t^6+O\left(t^7\right) \\
\end{array}
 \\\hline
\end{array}
\end{equation*}
    \caption{Evaluation of (\ref{HiggsBranchFormula}) for various values of $N$ and $N_f$ up to order $t^6$. The first line in each case corresponds to $G^{\circ} = \mathrm{SU}(N)$ and the second line to $G^{\circ} = \widetilde{\mathrm{SU}}(N)$. }
    \label{TableHSQCD}
\end{table}

In order to analyze the structure of the global symmetry, let us refine by adding fugacities for the global symmetry fugacities, according to Table \ref{tableSummaryInvariantsNf}. In all cases except when $N$ and $N_f$ are odd, the fundamental representation of the global symmetry has dimension $N_f$, and we can refine (\ref{HiggsBranchFormula}) as follows:
\begin{equation}
\label{HiggsBranchFormulaRefined1}
   HS_{(N,\,N_f)} = \int_{G^{\circ}} \d \eta_{G^{\circ}} (X) \frac{\mathrm{det} \left( 1 - t^2 \Phi_{\mathrm{Adj}} (X)\right)}{\mathrm{det} \left( 1 - t [1,0,\dots , 0]_{\mathfrak{g}} \times  \Phi_{\mathrm{F\bar{F}}} (X)\right) } \,  ,
\end{equation}
where $[1,0,\dots , 0]_{\mathfrak{g}}$ denotes the character of the fundamental of the global symmetry. In the case where $N$ and $N_f$ are odd, the fundamental has dimension $N_f -1$, and we consider instead 
\begin{equation}
\label{HiggsBranchFormulaRefined2}
   HS_{(N \textrm{ odd},\,N_f\textrm{ odd})} = \int_{\widetilde{\mathrm{SU}}(N)} \d \eta_{\widetilde{\mathrm{SU}}(N)} (X) \frac{\mathrm{det} \left( 1 - t^2 \Phi_{\mathrm{Adj}} (X)\right)}{\mathrm{det} \left( 1 - t (1+[1,0,\dots , 0]_{\mathfrak{g}}) \times  \Phi_{\mathrm{F\bar{F}}} (X)\right) } \,  . 
\end{equation}
Note that it is not guaranteed \textit{a priori} that the coefficients of the series (\ref{HiggsBranchFormulaRefined1}) and (\ref{HiggsBranchFormulaRefined2}) be characters with positive and integer coefficients, because of the pre-factor $\frac{1}{2}$ in the integration formula (\ref{eq:integration}). In particular, if one tries to insert $\mathrm{U}(N_f)$ characters in the above expressions, fractional coefficients inevitably appear. 
Note that it might look weird that the quarks don't fit in the fundamental representation of the flavor group and that an additional singlet be needed. However, looking back at where the structure of the global symmetry comes from, see equation (\ref{symmetryB}), this is related to the fact that an antisymmetric matrix of odd dimension always has at least one vanishing eigenvalue. 

In the notation that $[x_1,\cdots,\,x_n]_{\mathfrak{g}}$ represents the character of the representation with given Dynkin labels of the algebra $\mathfrak{g}$ we find the following decompositions: 
 \begin{eqnarray}
        HS_{(3,\,6)}&=&1+[2,0,0]_{C_3}\,t^2+\Big([0,0,1]_{C_3}+[1,0,0]_{C_3}\Big)\,t^3\nonumber\\ && +\Big(2\,[0,1,0]_{C_3}+2\,[0,2,0]_{C_3}+[4,0,0]_{C_3}+2\Big)\,t^4+O(t^5)\, .
      \end{eqnarray}
 \begin{eqnarray}
         HS_{(3,\,7)}&=&1+\Big([1,0,0]_{C_3}+[2,0,0]_{C_3}+1\Big)\,t^2+\Big([0,0,1]_{C_3}+[0,1,0]_{C_3}+[1,0,0]_{C_3}+1\Big)t^3 \nonumber \\ && +\Big([0,0,1]_{C_3} +2\,[0,1,0]_{C_3}+2\,[0,2,0]_{C_3}+3\,[1,0,0]_{C_3}+2\,[1,1,0]_{C_3}\nonumber \\ && +3\,[2,0,0]_{C_3} +[3,0,0]_{C_3} +[4,0,0]_{C_3}+3)\,t^4+O(t^5)\,.
      \end{eqnarray}
 \begin{eqnarray}
        HS_{(3,\,8)}&=&1+[2,0,0,0]_{C_4}\,t^2+\Big([0,0,1,0]_{C_4}+[1,0,0,0]_{C_4}\Big)\,t^3\\ \nonumber  && + \Big([0,0,0,1]_{C_4}+2\,[0,1,0,0]_{C_4}+2\,[0,2,0,0]_{C_4}+[4,0,0,0]_{C_4}+2\Big)\,t^4+O(t^5)\,.
      \end{eqnarray}
 \begin{eqnarray}
        HS_{(3,\,9)}&=&1+\Big([1,0,0,0]_{C_4}+[2,0,0,0]_{C_4}+1\Big)\,t^2  \\ \nonumber && +\Big([0,0,1,0]_{C_4}+[0,1,0,0]_{C_4}+[1,0,0,0]_{C_4}+1\Big)\,t^3 \nonumber \\ \nonumber &&  + \Big([0,0,0,1]_{C_4}+[0,0,1,0]_{C_4}+2\,[0,1,0,0]_{C_4}+2\,[0,2,0,0]_{C_4} +3\,[1,0,0,0]_{C_4}\\  && +2\,[1,1,0,0]_{C_4} +3
        \,[2,0,0,0]_{C_4}+[3,0,0,0]_{C_4}+[4,0,0,0]_{C_4}+3\Big)\,t^4+O(t^5)\, . \nonumber
      \end{eqnarray}
 \begin{eqnarray}
        HS_{(4,\,8)}&=&1+[0,1,0,0]_{D_4}\,t^2+\Big(2\,[0,0,0,2]_{D_4}+2\,[0,0,2,0]_{D_4}+2\,[0,2,0,0]_{D_4}+2\,[2,0,0,0]_{D_4}\nonumber \\ && +[4,0,0,0]_{D_4}+2\Big)\,t^4+O(t^5)\, .
      \end{eqnarray}
 \begin{eqnarray}
        HS_{(4,\,9)}&=&1+[0,1,0,0]_{B_4}\,t^2+\Big(2\,[0,0,0,2]_{B_4}+2\,[0,2,0,0]_{B_4}+2\,[2,0,0,0]_{B_4}\nonumber \\ && +[4,0,0,0]_{B_4}+2\Big)\,t^4+O(t^5)\, .
    \end{eqnarray}
In turn, for $HS_{(N,\,N_f)}^+$, the global symmetry is always $A_{N_f-1}$. Its character expansion expansion can be easily recovered from from the Highest Weights Generating function (HWG) \cite{Hanany:2014dia}, which reads \footnote{This expression is quite closed, even if not equal, to the expression of the HWG for a theory with gauge group $\mathrm{U}(N)$ and $2N_f$ flavors, that was already computed in \cite{Hanany:2016gbz} (see table 6 of \cite{Hanany:2016gbz} for its explicit expression).}
\begin{equation}
\label{genericHWG}
\textrm{HWG}^+_{(N,\,N_f)}=\textrm{PE}\left[t^2+\sum_{i=1}^{N-1}t^{2i}\mu_i\mu_{N_f-i}+t^{N}(\mu_{N}+\mu_{N_f-N})\right] \, ,
\end{equation}
where the various $\{\mu_i\}$ denote $\mathrm{SU}(N_f)$ highest weights.

From the above results it is natural to conjecture that the global symmetry is, at least, $\mathrm{SO}(N_f)$ for $\widetilde{\mathrm{SU}}(2N)$ and $\mathrm{Sp}(\lfloor\frac{N_f}{2}\rfloor)$ for $\widetilde{\mathrm{SU}}(2N+1)$. We summarize this in Figure \ref{figSummaryGlobalSymQCD}. This pattern ties in nicely with our conclusions from section \ref{sectionGtilde}. Note also that, besides the obvious loophole that strictly speaking we have expanded our Hilbert series in characters only up to a finite order, it may be that the true global symmetry is a larger group admitting $\mathrm{Sp}(\lfloor\frac{N_f}{2}\rfloor)$ or $\mathrm{SO}(N_f)$ as a subgroup.

\section*{Acknowledgements}
We would like to thank Elli Pomoni and Thomas Bourton for useful conversations. D.R-G and A.B. acknowledge support from the EU CIG grant UE-14-GT5LD2013-618459, the Asturias Government grant FC-15-GRUPIN14-108 and Spanish Government grant MINECO-16-FPA2015-63667-P. A.P. is supported by the German Research Foundation (DFG) via the Emmy Noether program “Exact results in Gauge theories". 

\appendix

\section{Invariant Theory}
\label{AppendixInvariant}

\subsection{Properties of invariant rings}

In this appendix, we review some salient aspects of the theory of invariants for a reductive group $G$, with the aim of proving the various statements made in section \ref{sectionGtilde}. A good reference on this topic is \cite{sturmfels2008algorithms}. 

Concretely, given a group $G$ and a linear $n$-dimensional representation $V$, we have by definition an action of $G$ on $V$. This action translates into an action of $G$ on the polynomial ring $\mathbb{C}[x_1 , \dots , x_n]:=\mathbb{C}[\mathbf{x}]$ where $(x_1 , \dots , x_n)$ are coordinates on $V$. The fundamental problem of invariant theory is then to describe the subring of invariant polynomials, denoted $\mathbb{C}[\mathbf{x}]^G$. 

Fortunately, by a theorem of Hilbert, $\mathbb{C}[\mathbf{x}]^G$ turns out to always be finitely generated as an algebra. This means that there exist $I_1 , \dots , I_m \in \mathbb{C}[\mathbf{x}]$ such that $\mathbb{C}[\mathbf{x}]^G$ consists exactly of polynomials of the $\mathbb{C}[\mathbf{I}]:=\mathbb{C}[I_1 , \dots , I_m]$. 

In some particular case, any $f \in \mathbb{C}[\mathbf{x}]^G$ can be written in a \emph{unique} way as a polynomial in the $I_1 , \dots , I_m$. When this is the case, the situation is particularly simple, since we have an isomorphism 
\begin{equation}
\label{freeGenerated}
    \mathbb{C}[\mathbf{x}]^G  \cong \mathbb{C}[\mathbf{I}] \, , 
\end{equation}
and we say that $\mathbb{C}[\mathbf{x}]^G$ is \emph{freely generated}. 

In general the situation is not so simple. However, it is not far from that, and only a slight modification of (\ref{freeGenerated}) is needed; namely, there always exist invariants $I_1 , \dots , I_m \in \mathbb{C}[\mathbf{x}]$ such that 
\begin{equation}
\label{Hironaka}
    \mathbb{C}[\mathbf{x}]^G \cong \bigoplus\limits_{j=1}^p J_j \mathbb{C}[\mathbf{I}] 
\end{equation}
where $J_1 , \dots , J_p \in \mathbb{C}[\mathbf{x}]$ are other invariants (and we can always set $J_1=1$). This is the so-called Cohen-Macaulay property of invariant rings. It is traditional to call the $I_1 , \dots , I_m$ the primary invariants, and $J_1 , \dots , J_p$ the secondary invariants.\footnote{Intuitively, the secondary invariants can be considered as ``exceptional" ones, in the sense that higher powers of them are not necessary to generate the invariants -- this is what (\ref{Hironaka}) says. This can be rephrased saying that they ``satisfy relations". }
Once the decomposition (\ref{Hironaka}), called the Hironaka decomposition, is known, one can compute the Hilbert series of $\mathbb{C}[\mathbf{x}]^G$: 
\begin{equation}
\label{HShironaka}
    HS(\mathbb{C}[\mathbf{x}]^G , t) = \frac{\sum\limits_{j=1}^p t^{\deg J_j}}{\prod\limits_{i=1}^m (1-t^{\deg I_i})} \, . 
\end{equation}
The denominator captures the primary invariants, while the numerator is concerned with the secondary invariants. 
As a particular case, when $\mathbb{C}[\mathbf{x}]^G$ is freely generated, the Hironaka decomposition reduces to (\ref{freeGenerated}) -- there is only one secondary invariant, the trivial $1$ -- and the Hilbert series (\ref{HShironaka}) becomes
\begin{equation}
    \textrm{Freely generated :   } HS(\mathbb{C}[\mathbf{x}]^G , t) = \frac{1}{\prod\limits_{i=1}^m (1-t^{\deg I_i})} \, . 
\end{equation}

The plethystic logarithm of a function $f(t)$ such that $f(0)=1$ is defined as  
\begin{equation} 
    \mathrm{PL}[f(t)] = \sum\limits_{k=1}^{\infty} \frac{\mu (k)}{k} \log f(t^k) \, ,  
\end{equation}
where $\mu(k)$ denotes the M\"obius function. As the usual logarithm, it transforms products into sums, $\mathrm{PL}[fg] = \mathrm{PL}[f] + \mathrm{PL}[g]$, and we have $\mathrm{PL}[1-t^a]=-t^a$ for any positive integer $a$. From these two properties, we deduce that the plethystic logarithm of the Hilbert function of a freely generated ring is a (finite) polynomial with positive integer coefficients. 

\subsection{Algorithm to find invariants}
\label{sectionAlgoInv}

In this subsection, we recall an algorithm that can be used to compute the ring of invariants of a group $G$ in a linear representation $V$. It is based on two tools from commutative algebra: 
\begin{enumerate}
    \item An averaging operator over the group $G$, which sends any element of $\mathbb{C}[\mathbf{x}]$ to an invariant in $\mathbb{C}[\mathbf{x}]^G$. For the principal extensions, this averaging operator is for practical computations nothing but the integration formula from section \ref{sectionIntegrationFormula}. 
    \item An algorithm to compute the Hilbert series of the polynomial ring $\mathbb{C}[\mathbf{I}]$ where the $I_1 , \dots , I_m$ are elements of $\mathbb{C}[\mathbf{x}]$. Such algorithms are known (see for instance \cite{cox2007ideals}). 
\end{enumerate}
Finally, we need Molien's theorem, which says that the Hilbert series of the ring of invariants of $G$ on $V$ is the average of $(\det (1-t \Phi_{V} (X)))^{-1}$ for $X \in G$, 
\begin{equation}
\label{MolienIntegral}
   \mathcal{M}_{G,V} (t) = \int_G \frac{\d \eta_G (X)}{\det (1-t \Phi_{V} (X))} \, . 
\end{equation}

Now, the algorithm is based on the following remark: if $I_1 , \dots , I_m$ are invariants of $G$, then obviously we have the inclusion $\mathbb{C}[\mathbf{I}] \subset \mathbb{C}[\mathbf{x}]^G$. We can compute the Hilbert series of both rings using point 2. above for the first one, and Molien's formula and the averaging operator for the second one. If they agree, then the subspaces consisting of degree $k$ polynomials have the same dimension, and therefore are equal, for all $k$, which means that $\mathbb{C}[\mathbf{I}] = \mathbb{C}[\mathbf{x}]^G$. If the two Hilbert series don't agree, one looks at the smallest degree $k$ where there is a difference, and knows that an invariant of that degree must be found -- and an easy way to generate invariants is to apply the averaging operator to arbitrary degree $k$ polynomials. 

Note that using the elimination ideal, one also has access to the complete list of relation that defines the invariant ring -- by definition, this list is empty if and only if the ring is freely generated. 

\subsection{\texorpdfstring{Invariants in some representations of $\widetilde{\mathrm{SU}}(N)$}{Invariants in some representations}}
\label{sectionInvBifund}

\subsubsection*{Invariants in the adjoint}

As a first illustrations of the concepts of section \ref{sectionAlgoInv}, we look at the invariants of $\widetilde{\mathrm{SU}}(N)$ in the adjoint representation. As already mentioned, the averaging operator in this case is the Weyl integration formula, and the Molien integral (\ref{MolienIntegral}) coincides with the Coulomb branch Hilbert series (\ref{eq:cindex}). As explained there, this coincides manifestly with the Hilbert series generated by the primary and secondary invariants of section \ref{sectionRepandInv}, and therefore constitutes a proof that these invariants indeed generate the full ring.

\subsubsection*{Invariants in the bifundamental}

One can compute, by direct evaluation using (\ref{eq:integration}) that the Molien integral in the bifundamental is 
\begin{equation}
\label{MolienFF}
 \mathcal{M}_{\widetilde{\mathrm{SU}}(N),\textrm{F}\bar{\textrm{F}}} (t) =
    \begin{cases}
    \frac{1}{1-t^2} & \textrm{ for } N \textrm{ odd} \\
    \frac{1}{1-t^4} & \textrm{ for } N \textrm{ even}
    \end{cases}
\end{equation}
We have seen in section \ref{sectionRepandInv} that for $N$ odd, an invariant is $y^T x$ and for $N$ even an invariant is $(y^T x)^2$. Clearly, in both cases the Hilbert series of the associated ring coincides with the result (\ref{MolienFF}), and this proves that this is a correct description of the ring of invariants of $\widetilde{\mathrm{SU}}(N)$ in the bifundamental representation. 

\subsubsection*{Invariants in $N_f$ bifundamentals}

To illustrate the fact that the invariants quickly become intricate when the number of flavors $N_f$ becomes large, we report the result for $N = N_f = 3$. 

In that case, the $\mathrm{SU}(3)$ situation is relatively simple: 
\begin{equation}
   \mathcal{M}_{\mathrm{SU}(3),3 \otimes \textrm{F}\bar{\textrm{F}}} (t) = \textrm{PE}[9 t^2 + 2 t^3 - t^6] \, .
\end{equation}
The ring is a complete intersection generated by the $N_f^2$ mesons and the two baryons, with one relation. But going to $\widetilde{\mathrm{SU}}(3)$, we obtain 
\begin{equation}
\label{complicatedExample}
   \mathcal{M}_{\widetilde{\mathrm{SU}}(3),3 \otimes \textrm{F}\bar{\textrm{F}}} (t) = \frac{1+3 t^4+3 t^5+t^9}{\left(1-t^2\right)^6 \left(1-t^3\right) \left(1-t^4\right)^3} \, . 
\end{equation}
As expected, this is of the form (\ref{HShironaka}), but the number of primary and secondary invariants is very different than in the $\mathrm{SU}(3)$ case above.

\section{Technical Details}
\label{AppendixTechnical}

In this section, we clarify the relation between complex conjugation, that we will denote by $\mathcal{C}$, and the operation $\mathcal{P}$ described in the text, in the algebra $\mathfrak{g} = \mathfrak{su}(N)$ of the group $G=\mathrm{SU}(N)$. We call $\mathrm{Aut} (\mathfrak{g})$ the group of Lie algebra automorphisms of $\mathfrak{g}$. The group $\mathrm{Int} (\mathfrak{g})$ of inner automorphisms is the subgroup of $\mathrm{Aut} (\mathfrak{g})$ of Lie algebra automorphisms of the form $M \rightarrow g M g^{-1}$ for $g \in G$.\footnote{The definition we give here works because $\mathrm{SU}(N)$ is a closed group of complex matrices. In general, if $\mathfrak{g}$ is a finite-dimensional Lie algebra over $\mathbb{R}$, $\mathrm{Int} (\mathfrak{g})$ is defined as the analytic subgroup of the $\mathbb{R}$-linear automorphisms of $\mathfrak{g}$ whith Lie algebra $\mathrm{ad} \, \mathfrak{g}$. }  One can show (see Threorem 7.8 in \cite{knapp2013lie}) that $\mathrm{Aut} (\mathfrak{g}) / \mathrm{Int} (\mathfrak{g})$ is isomorphic to the group of automorphisms of the Dynkin diagram of $\mathfrak{g}$, which is $\{1 , \mathcal{P}\}$. 

We now give a very explicit description of the automorphism $\mathcal{P} : \mathfrak{g} \rightarrow \mathfrak{g}$. For that, we first go to the complexification of $\mathfrak{g}$, which is the algebra $\mathfrak{g}^{\mathbb{C}} = \mathfrak{sl}(N , \mathbb{C})$ of all traceless $N \times N$ matrices with complex coefficients. Let us call $E_{i,j}$ the matrix where the only non-vanishing coefficient is a $1$ at position $(i,j)$. The Cartan subalgebra of diagonal matrices is generated by the $H_{i} = E_{i,i} - E_{i+1,i+1}$ with $i=1 , \dots N-1$. The non-diagonal part is generated by the $E_{i,j}$ with $i \neq j$. One can show that the action of $\mathcal{P} : \mathfrak{g}^{\mathbb{C}} \rightarrow \mathfrak{g}^{\mathbb{C}}$ on these generators is\footnote{This can be shown by recursion on $|i-j|$. } 
\begin{equation}
\label{actionPCartan}
    \mathcal{P} (H_i) = H_{N-i} \, . 
\end{equation}
\begin{equation}
\label{actionPCcomplexifiedAlg}
    \mathcal{P} (E_{i,j}) = (-1)^{1+i-j} E_{N+1-j , N+1-i} \, . 
\end{equation}
The action on $\mathfrak{g}$ follows by linearity:
\begin{equation}
    \mathcal{P} (E_{i,j}+E_{j,i}) = (-1)^{1+i-j} ( E_{N+1-j , N+1-i} + E_{N+1-i , N+1-j}) \, ,
\end{equation}
\begin{equation}
    \mathcal{P} (i(E_{i,j}-E_{j,i})) = (-1)^{1+i-j} i ( E_{N+1-j , N+1-i} - E_{N+1-i , N+1-j}) \, .
\end{equation}
This is summarized as follows: the action of $\mathcal{P}$ on the matrix $X = (x_{i,j})_{1 \leq i,j \leq N}$ is given by 
\begin{equation}
\label{actionPMatrixComponents}
    x \mapsto x^{\mathcal{P}} = \left( (-1)^{1+i-j} x_{N+1-j,N+1-i}\right)_{1 \leq i,j \leq N} \, . 
\end{equation}
In particular, we find by direct evaluation that 
\begin{eqnarray}
\label{proofTr}
     \mathrm{Tr} \left( (x^{\mathcal{P}})^k  \right) &=& \sum\limits_{i_1 , \dots , i_k} (x^{\mathcal{P}})_{i_1 , i_2} \dots (x^{\mathcal{P}})_{i_k , i_1} \nonumber \\
     &=& \sum\limits_{i_1 , \dots , i_k} (-1)^{1+i_1-i_2} x_{N+1-i_2 , N+1-i_1} \dots (-1)^{1+i_k-i_1} x_{N+1-i_1 , N+1-i_k} \nonumber \\
      &=& \sum\limits_{i_1 , \dots , i_k} (-1)^{k} x_{N+1-i_2 , N+1-i_1} \dots  x _{N+1-i_1 , N+1-i_k} \nonumber \\
          &=& (-1)^{k} \sum\limits_{i_1 , \dots , i_k}  x_{i_2 , i_1} \dots  x _{i_1 , i_k}  \\ 
          &=& (-1)^{k} \, \mathrm{Tr} \left( (x^{T})^k \right) \nonumber \\ 
          &=& (-1)^{k} \, \mathrm{Tr} \left( x^k \right) \, .  \nonumber 
\end{eqnarray}

The complex conjugation is the map 
\begin{eqnarray}
    \mathcal{C} &:& \mathfrak{g} \rightarrow \mathfrak{g} \\
    & & M \mapsto  - M^{\ast} = - M^T\, . \nonumber
\end{eqnarray}
The minus sign is a consequence of the choice (\ref{definitionsuN}).\footnote{Indeed, if we set $\mathcal{C} (M) = \alpha M^{\ast}$, then the fact that $\mathcal{C}$ preserves the Lie bracket $i[\cdot , \cdot]$ becomes $\mathcal{C}(i[M,M']) = -i \alpha [M,M']^{\ast} = i \alpha^2  [M,M']^{\ast}$, so $\alpha = -1$. } We now show that $\mathcal{P}$ and $\mathcal{C}$ belong to the same class in $\mathrm{Aut} (\mathfrak{g}) / \mathrm{Int} (\mathfrak{g})$. For that, it is sufficient to find a matrix $A$ such that for all $M \in \mathfrak{g}$, 
\begin{equation}
\label{equationConjugation}
    \mathcal{P}(M) = A \mathcal{C}(M) A^{-1} \, . 
\end{equation}
One can check that the matrix 
\begin{equation}
\label{matrixA}
    A = \left( 
    \begin{array}{cccccc}
         & & & & & 1 \\
         & & & & -1 &  \\
         & & & 1 & &  \\
         & & \dots & & &   \\
         & \dots& & & &   \\
         (-1)^{N-1} & & & & &   \\
    \end{array}
    \right) \, ,
\end{equation}
satisfies equation (\ref{equationConjugation}). This means that both $\mathcal{P}$ and $\mathcal{C}$ are outer automorphisms. However, they are very different operators, as can be seen from their spectrum of eigenvalues. The algebra $\mathfrak{g}$ is the direct sum of the space of real symmetric traceless matrices, of dimension $\frac{1}{2}N(N+1)-1$, and purely imaginary antisymmetric matrices, of dimension $\frac{1}{2}N(N-1)$. So the spectrum of $\mathcal{C}$ is 
\begin{equation}
\label{spectrumC}
    \mathrm{Spec}(\mathcal{C}) = \left\{ (+)^{\frac{1}{2}N(N-1)} , (-)^{\frac{1}{2}N(N+1)-1} \right\} \, . 
\end{equation}
On the other hand, the operator $\mathcal{P}$ leaves invariant a space of dimension $N+(-1)^N$ on which is acts as $-1$, and induces a transformation of type 
\begin{equation}
    \left( 
    \begin{array}{cc}
        0 & 1 \\
        1 & 0
    \end{array}
    \right)
\end{equation}
on $ \left[ \frac{N-1}{2}\right] \left( 2\left[ \frac{N-1}{2}\right] + 2+(-1)^N\right) = \frac{1}{2} \left((N-1) N-(-1)^N-1\right)$ spaces of dimension $2$. In total, we then have a spectrum 
\begin{equation}
    \mathrm{Spec}(\mathcal{P}) = \left\{ (+)^{\frac{1}{2} \left(N(N-1) -(-1)^N-1\right)} ,  (-)^{\frac{1}{2} \left(N(N+1) +(-1)^N-1\right)} \right\} \, . 
\end{equation}
Clearly, the spectra are different in general. However, note that when $N$ is odd this is equal to (\ref{spectrumC}).

\section{\texorpdfstring{The Coulomb branch of $\widetilde{E}_6$ theory}{The Coulomb branch of E6 theory}}
\label{E6theory}

For completeness, in this appendix we focus on the Coulomb branch of the $\widetilde{E_6}$ QFT. The corresponding index reads 
\begin{equation}
\mathcal{I}_{\widetilde{E_6}}^{\mathrm{Coulomb}}(t)=    \int_{\widetilde{E_6}} \d \eta_{\widetilde{E_{6}}}(X) \frac{1}{\det(1-t\Phi_{\mathrm{Adj}}^{\widetilde{E_6}}(X))} \, .
\end{equation}
The integration over the connected part of the group can be performed analytically and it reads \cite{Gadde:2011uv} 
\begin{equation}
I_{E_6}^{\mathrm{Coulomb}}(t) = \textrm{PE}\left[\sum_{i \in J} t^i \right] \, ,
\end{equation}
where $\mathcal{J} =\{2,5,6,8,9,12 \}$ is the set of the degrees of fundamental invariants. On the other hand the integration over the non-connected component can be performed through a brute force computation using Mathematica up to an enough high order of the expansion. Putting together the contributions arising from the two components is natural to conjecture that the full Hilbert series is given by
\footnote{This expression coincides with the Coulomb branch Hilbert series of one of the QFTs discussed in \cite{Tom}.}
 \begin{equation}
 \label{eq:indexe6}
     \mathcal{I}^{\textrm{Coulomb}}_{\widetilde{E_6}}(t) = \frac{1}{2}\left[\ \prod\limits_{i \in \mathcal{J}}(1-t^i)^{-1} + \prod\limits_{i \in \mathcal{J}}(1-(-t)^i)^{-1} \right] \, , \end{equation}
 where as before $\mathcal{J} = \{2,5,6,8,9,12 \}$. Remarkably, as it as also discussed in \cite{Tom}, the corresponding Coulomb branch is not freely generated since the PL of the Hilbert series (\ref{eq:indexe6}) reads
 \begin{equation}
   \textrm{PL}\left[\mathcal{I}^{\textrm{Coulomb}}_{\widetilde{E_6}}(t)\right] = t^2+t^6+t^8+t^{10} + t^{12}+t^{14}+t^{18}-t^{28} \, . 
 \end{equation}

\bibliographystyle{JHEP}
\bibliography{bibli.bib}

\end{document}